\documentclass[prd,preprint,tightenlines,floatfix,showpacs,preprintnumbers,nofootinbib,eqsecnum]{revtex4}

 \usepackage[dvips,final]{graphicx}
  \usepackage{amssymb}
   \usepackage{amsmath}
    \usepackage{amsfonts}
     \usepackage{epsfig}
      \usepackage{bm}% bold math

\usepackage[section]{placeins}

\usepackage{multirow}
\usepackage{ctable}
\usepackage{booktabs}
\usepackage{array}
\usepackage{tabularx}
\usepackage{xcolor}
\usepackage{pstricks}
\def\Pom{{\bf I\!P}}

\begin{document}

\vfill
\title{
Double-scattering mechanism in
exclusive $A A \to A A \rho^0 \rho^0$ reaction
at ultrarelativistic collisions}

\author{Mariola K{\l}usek-Gawenda}
\email{mariola.klusek@ifj.edu.pl} \affiliation{Institute of Nuclear
Physics PAN, PL-31-342 Cracow,
Poland}
\author{Antoni Szczurek}
\email{antoni.szczurek@ifj.edu.pl} \affiliation{Institute of Nuclear
Physics PAN, PL-31-342 Cracow,
Poland and\\
University of Rzesz\'ow, PL-35-959 Rzesz\'ow, Poland}

\date{\today}

\begin{abstract}
We calculate, for the first time, differential distributions for double $\rho^0$ meson
production in exclusive ultraperipheral, ultrarelativistic collisions
via a double scattering mechanism. The calculations are done in impact
parameter space. The cross section for $\gamma A \to \rho^0 A$ is
parametrized based on an existing calculation. Smearing of $\rho^0$ masses
is taken into account. The results of our calculations are compared at the RHIC
energy to the contribution of the two-photon mechanism discussed
previously in the literature. 
The cross section for the double scattering mechanism
is found to be an order of magnitude larger at $M_{\rho \rho} <$ 2 GeV
and more than two orders of magnitude at $M_{\rho \rho} >$ 3 GeV, 
than that for the photon-photon mechanism.
Compared to the two-photon mechanism
the double scattering mechanism populates somewhat larger $\rho^0 \rho^0$ 
invariant masses and larger rapidity distances between the two $\rho^0$
mesons and gives a
significant contribution to the $A A \to A A \pi^+ \pi^- \pi^+ \pi^-$
reaction.
Some observables related to charged pions are presented too.
We compare the results of our calculation with the STAR collaboration
results  on four charged pion production.
While the shape in invariant mass of the four-pion system is very
similar to the measured one, 
the predicted cross section constitutes only 20 \% of the measured one.
We discuss a possibility of identifying 
the double scattering mechanism at the LHC.
\end{abstract}

\pacs{25.75.Dw -Particle production (relativistic collisions)\\
      13.25.-k -Deacy mesons hadronic }

\maketitle

%----------------------------
\section{Introduction}
%----------------------------

The exclusive production of simple final states
in ultraperipheral, ultrarelativistic heavy ion collisions
is a special class of nuclear reactions \cite{reviews}.
At high energies and due to large charges of colliding nuclei there are 
two categories of the underlying reaction mechanisms.
One is the photon-photon fusion and the second is photoproduction
(photon fluctuation into hadronic or quark-antiquark components
and its transformation into a simple final state). 
The competition between these two mechanisms was studied only for few
cases. 

For $\rho^0 \rho^0$ production only the photon-photon
mechanism was discussed in the literature 
\cite{GM2003,GMS2006,KSS2009}.
In Ref. \cite{KSS2009} we have made a first realistic estimate
of the corresponding cross section.

The cross section for single meson production was predicted to be large
\cite{KN1999,FSZ2003,GM2005}. Measurements at RHIC confirmed
the size of the cross section \cite{STAR2008_rho0} at midrapidities,
but were not able to distinguish between different models that predicted
different behaviour at large (unmeasured) (pseudo)rapidities.
The large cross section for single $\rho^0$ production suggests
that the cross section for double scattering process should be also rather
large. The best example of a similar type of reaction is the production
of $c \bar c c \bar c$ final state in proton-proton collisions 
which was measured recently by the LHCb
collaboration \cite{LHCb_ccbarccbar}. This was predicted and explained 
in Ref. \cite{ccbarccbar} as a double-parton scattering effect.
There, the cross section for the $c \bar c c \bar c$ final state is
of the same order of magnitude as the cross section for single $c \bar c$
pair production.
The situation for exclusive $\rho^0$ production is somewhat similar.
Due to easier control of absorption effect, the impact parameter 
formulation seems in the latter case the best approach.

In the present paper, we wish to focuss on double $\rho^0$ production in 
ulraperipheral, ultrarelativistic heavy ion reactions.
We wish to study differential single particle distributions for 
the $\rho^0$ mesons, as well as correlations between the $\rho$ mesons,
also for the photon-photon component. 
A comparison to the results for photon-photon process will be done too, 
in order to understand how to identify the double photoproduction process. 
We intend to take into account the decay of $\rho^0$ mesons 
into charged pions, in order to take into account some
experimental cuts of existing experiments.
We shall discuss how to identify the double-scattering mechanism
at the LHC.

%-----------------------
\section{Formalism}
%-----------------------

%-------------------------------------------------------
\subsection{Single scattering mechanisms}
%-------------------------------------------------------

\begin{figure}[!h]
\includegraphics[width=6.0cm]{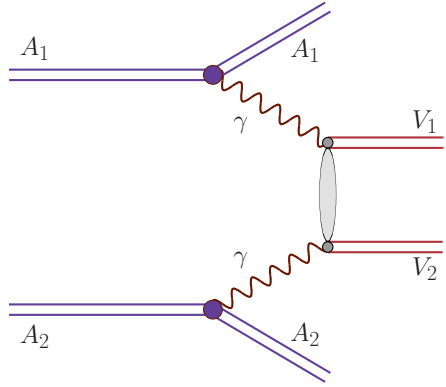}
   \caption{
\small The photon-photon mechanism of two vector meson
production in ultrarelativistic ultraperipheral collisions.}
\label{fig:photon_photon}
\end{figure}

Most of the previous analyses in the literature concentrated on 
production of pairs of mesons in photon-photon processes 
(see Fig. \ref{fig:photon_photon}).
In the past we have studied both exclusive $\rho^0 \rho^0$ productions 
\cite{KSS2009} and recently exclusive production of $J/\psi J/\psi$ pairs.

%-----------------------------------------------------------------------------
%------------------------------------------------------------------------------

In the case of double $\rho^0$ production there are two mechanisms.
At larger photon-photon energies the pomeron/reggeon exchange mechanism
is the dominant one, while close to $\rho^0 \rho^0$ threshold a large 
enhancement was observed. The latter mechanism is not well
understood.
In Ref. \cite{KSS2009} this enhancement of the cross section was
parametrized. In the present paper we shall concentrate rather on larger
photon-photon energies (larger dimeson invariant masses).

In the case of double $J/\psi$ production there are also two distinct mechanisms.
At low photon-photon energies a box mechanism dominates, and at larger
energies the two-gluon exchange mechanism takes over.
 
The elementary cross section for $\gamma \gamma \to \rho^0 \rho^0$
has been measured in the past \cite{rho0rho0_data} for not too large energies.
The measured cross section shows a characteristic bump at about
$M_{\rho \rho} \sim$ 1.5 GeV. The origin of this bump is so far not
well understood.
In Ref. \cite{KSS2009} we have parametrized the measured cross 
section and used the resulting elementary cross section as the input 
for nuclear calculations.

%In Fig. \ref{fig:possible_mechanisms} we show 
There are many potential mechanisms contributing to the bump. 
Since the calculation of the corresponding cross sections is not always easy, 
we leave theoretical studies
of the underlying dynamics for future studies. At somewhat larger photon-photon
energies ($\rho^0 \rho^0$ invariant masses in the case of nuclear collisions)
another mechanism, which can be relatively well calculated, plays the
dominant role, namely soft virtual (quasi-real) vector meson rescattering.
The corresponding matrix element for not too large $\rho^0$ meson
transverse momenta can be parametrized in the VDM-Regge language 
\cite{KSS2009}. 
At large transverse momenta of the $\rho^0$ meson, two-gluon exchange
should become important (see a discussion of two-gluon exchange for 
$J/\Psi J/\Psi$ production in Ref. \cite{BCKSS2013}). In the present analysis
we shall discuss only the soft scattering mechanisms.
The hard mechanism may be important at the LHC.
It is interesting if the two mechanisms can be identified by imposing
kinematical cuts, or by looking at special observables.

The nuclear cross section for the photon-photon mechanism is calculated
in the impact parameter space as:
\begin{eqnarray}
 \sigma\left( AA \to AA \rho^0\rho^0 \right) &=& 
\int \hat{\sigma} \left( \gamma\gamma \to \rho^0\rho^0 ; W_{\gamma\gamma}\right) 
S^2_{abs}\left( \bf{b} \right) N(\omega_1, {\bf b}_1) N(\omega_2, {\bf b}_2) \nonumber \\
& \times & d^2 {\bf b}_1 d^2 {\bf b}_2 d\omega_1 d\omega_2  \; , 
\end{eqnarray}
where
\begin{equation}
 S^2_{abs}\left( \bf{b} \right) = \theta ({\bf b}-2R_A) = 
\theta (|{\bf b}_1-{\bf b}_2|-2R_A)  \; .
\end{equation}
This can be written equivalently as:
\begin{eqnarray}
 \sigma\left( AA \to AA \rho^0\rho^0 \right) &=& 
\int \hat{\sigma} \left( \gamma\gamma \to \rho^0\rho^0 ; W_{\gamma\gamma}\right) 
S^2_{abs}\left( \bf{b} \right) N(\omega_1, {\bf b}_1) N(\omega_2, {\bf b}_2) \nonumber \\
& \times & \frac{W_{\gamma\gamma}}{2} d^2 {\bf b}_1 d^2 {\bf b}_2 d
W_{\gamma\gamma} d Y_{\rho^0\rho^0} \; .
\end{eqnarray}
Four-momenta of $\rho^0$ mesons in the $\rho^0\rho^0$ center of mass frame
can be written as:
\begin{equation}
 E_{\rho^0} = \frac{\sqrt{\hat{s}}}{2}  \; ,
\end{equation}
\begin{equation}
 p_{\rho^0} = \sqrt{\frac{\hat{s}}{4} - m_{\rho^0}^2} \; ,
\end{equation}
\begin{equation}
 p_{t,\rho^0} = \sqrt{1-z^2} p_{\rho^0} \; ,
\end{equation}
\begin{equation}
 p_{l,\rho^0} = z p_{\rho^0} \; .
\end{equation}
The rapidity of each of the $\rho^0$ mesons ($i$ = 1, 2) can be calculated as:
\begin{equation}
 y_{i} = Y_{\rho^0\rho^0} + y_{i/\rho^0\rho^0}(W,z) \; ,
\end{equation}
where $z$ is related to $\rho^0$ meson transverse momentum.
Other kinematical variables are calculated by adding relativistically 
velocities:
\begin{equation}
 \overrightarrow{v}_i = \overrightarrow{V}_{\rho^0\rho^0} \oplus
 \overrightarrow{v}_{i/\rho^0\rho^0}   \; ,
\end{equation}
\begin{equation}
 \overrightarrow{V}_{\rho^0\rho^0} =
 \frac{\overrightarrow{P}_{\rho^0\rho^0}}{E_{\rho^0\rho^0}} \; .
\end{equation}
The energies of photons can be expressed in terms of our integration
variables
\begin{equation}
 \omega_{1/2} = \frac{W_{\gamma\gamma}}{2} \exp (\pm Y_{\rho^0\rho^0})
\end{equation}
from the energy-momentum conservation:
\begin{eqnarray}
 E_{\rho^0\rho^0} &=& \omega_1 + \omega_2 \; , \nonumber \\
 P^z_{\rho^0\rho^0} &=& \omega_1 - \omega_2 \; .
\end{eqnarray}
The total elementary cross section can be calculated as:
\begin{equation}
 \hat{\sigma}(\gamma\gamma\to\rho^0\rho^0) =
 \int^{t_{max}(\hat{s})}_{t_{min}(\hat{s})}
 \frac{d\hat{\sigma}}{d\hat{t}} d\hat{t} \; ,
\end{equation}
where
\begin{equation}
 \frac{d\hat{\sigma}(\gamma\gamma\to\rho^0\rho^0)}{d\hat{t}} = \frac{1}{16\pi\hat{s}^2} |\mathcal{M}_{\gamma\gamma\to\rho^0\rho^0}|^2 \; .
\end{equation}
The high $W_{\gamma \gamma}$-(sub)energy matrix element is calculated
in a VDM-Regge approach \cite{KSS2009} as
\begin{eqnarray}
{\cal M}_{\gamma \gamma \to \rho^0 \rho^0} &=&
C_{\gamma \to \rho^0} C_{\gamma \to \rho^0} {\hat{s}}
\left( \eta_{\Pom}(\hat{s},\hat{t}) C_{\Pom} 
\left( \frac{\hat{s}}{s_0} \right) ^{\alpha_{\Pom}(t)-1}
+ \eta_{R}(\hat{s},\hat{t}) C_{R} 
\left( \frac{\hat{s}}{s_0} \right)^{\alpha_{R}(t)-1} \right) 
\nonumber \\
&\times& F(\hat{t}, q_1^2 \approx 0) F(\hat{t}, q_2^2 \approx 0) \; .
\label{VDM_Regge_amplitude}
\end{eqnarray}
This seems to be consistent with the existing world experimental data 
on total $\gamma \gamma \to \rho^0 \rho^0$ cross section \cite{KSS2009}.
The $C_{\gamma \to \rho^0}$ factors,
describing transformation of photons to (virtual) vector mesons, are 
calculated in the Vector Dominance Model (VDM). The parameters responsible
for energy dependence are taken from the Donnachie-Landshoff parametrization
of the total proton-proton and pion-proton cross sections \cite{DL92}
assuming Regge factorization. The slope parameter ($B$) is taken to be 
$B$ = 4 GeV$^{-2}$.
How the form factors $F(\hat{t}, q^2)$ are parametrized is described in detail 
in Ref.~\cite{KSS2009}. When calculating kinematical variables, a fixed
resonance position $m_{\rho} = m_R$ is taken for the 
$\gamma \gamma \to \rho^0 \rho^0$. Mass smearing could be included
if necessary. 

The differential distributions can be obtained by replacing total
elementary cross section by
\begin{equation}
 \hat{\sigma}(\gamma\gamma\to\rho^0\rho^0) = \int 
\frac{d\hat{\sigma}(\gamma\gamma\to\rho^0\rho^0)}{d {p_t}} d {p_t} \; ,
\end{equation}
where
\begin{equation}
  \frac{d\hat{\sigma}}{d {p_t}}  =  \frac{d\hat{\sigma}}{d {p_t}^2} \frac{d {p_t}^2}{d {p_t}} 
				 =  \frac{d\hat{\sigma}}{d {p_t}^2} 2 p_t 
				 =  \frac{d\hat{\sigma}}{d\hat{t}} 
                       | \partial \hat{t} / \partial p_t^2 | 2 p_t \; .
\label{variable_transformation}
\end{equation}
The first $\rho^0$ is emitted in the forward and the second $\rho^0$
in the backward direction in the $\gamma \gamma \to \rho^0 \rho^0$
center-of-mass system.
Finally the following three-dimensional maps (grids) are prepared separately
for the low-energy bump and VDM-Regge components:
\begin{equation}
\frac{d \sigma_{AA \to AA \rho^0 \rho^0}}{dy_1 dy_2 d p_t} \; .
\label{maps}
\end{equation}
The maps (grids) are used then to calculate distributions of pions from 
the decays of $\rho^0$ mesons produced in the photon-photon fusion.

%-----------------------------------------------------
\subsection{Single $\rho^0$ production}
%-----------------------------------------------------
The cross section for single vector meson production, differential
in impact factor and vector-meson rapidity, reads:
\begin{equation}
\frac{d \sigma}{d^2 b dy} = 
\omega_1 \frac{d {\tilde N}}{d^2 b d \omega_1} \sigma_{\gamma A_2 \to V
  A_2}(W_{\gamma A_2}) +
\omega_2 \frac{d {\tilde N}}{d^2 b d \omega_2} \sigma_{\gamma A_1 \to V
  A_1}(W_{\gamma A_1}) \; ,
\label{single_vector_meson}
\end{equation}
where $\omega_1 = m_{\rho^0}/2 \exp(+y)$ and $\omega_2 = m_{\rho^0}/2 \exp(-y)$.
Here the flux factor of equivalent photons, ${\tilde N}$, is in principle 
a function of heavy ion - heavy ion impact parameter $b$ and not of 
photon-nucleus impact parameter as is often done in the literature. 
The effective impact factor can be formally written as the convolution
of real photon flux in one of the nuclei and effective 
strength for interaction of the photon with the second nucleus
\begin{equation}
\frac{d {\tilde N}}{d^2 b d \omega} = \int \frac{d N}{d^2 b_1 d \omega}
\frac{S(b_2)}{\pi R_A^2} d^2 b_1 \approx 
\frac{dN}{d^2b d \omega}  \; ,
\end{equation}
where $\vec{b}_1 = \vec{b} + \vec{b}_2$ and $S(b_2) = \theta(R_A - b_2)$,
i.e., it is assumed that the collision occurs when the photon hits
the nucleus. For the photon flux in the second nucleus
one needs to replace 1 $\to$ 2 (and 2 $\to$ 1).
 
In general one can write:
\begin{equation}
\sigma_{\gamma A \to V A}(W) = 
\frac{d \sigma_{\gamma A \to V A}(W,t=0)}{dt} 
\int_{-\infty}^{t_{max}} dt |F_A(t)|^2 \; .
\label{gammaA_VA} 
\end{equation}
The second factor includes the $t$-dependence for the
$\gamma A \to V A$ subprocess which is due to coherent $q \bar q$ dipole
rescattering off a nucleus. To good approximation this
is dictated by the nuclear strong form factor.
In practical calculations, we approximate the nuclear strong form factor 
by the nuclear charge form factor.
The $t_{max}$ is calculated from kinematical dependences, $t_{max} = - (m_{\rho^0}^2/(2\omega_{lab}))^2$.
The first term in Eq. (\ref{gammaA_VA}) is usually weakly dependent
on the $\gamma A$ energy. For the $\rho^0$ meson it is practically
a constant \cite{KN1999}:
\begin{equation}
\frac{d\sigma(\gamma+A \to \rho^0 A;t=0)}{dt} 
\approx \mbox{const} \; .
\label{gammaA_rho0A}
\end{equation}
In the present exploratory calculation 
the constant is taken to be (see \cite{KN1999})
420 mb/GeV$^2$ for RHIC and 450 mb/GeV$^2$ for LHC.
These are cross sections for $W_{\gamma p}$ energies relevant 
for midrapidities at $\sqrt{s_{NN}} =$ 200 GeV and 5.5 TeV,
respectively.

A more refined treatment will be presented elsewhere when discussing
different models of $\rho^0$ photoproduction.
The second term depends on $t_{max}$ which in turn depends rather on 
running $\rho^0$ meson mass than on resonance position.

The cross section for the $\gamma A \to V A$ reaction could be also
calculated e.g., in the QCD dipole picture in a (convenient) so-called 
mixed representation (see e.g., \cite{GM2006,Lappi2013}). 
Slightely more complicated momentum space formulation of the vector meson
production on nuclei was discussed in Ref. \cite{CSS2012}.

At high energy the imaginary part of the amplitude for the $\gamma A \to V A$
process can be written as \cite{KNNZ,NNZ}:
\begin{equation}
\Im \left( A_{\gamma A \to V A}(W) \right) =
\Sigma_{\lambda \bar \lambda} \int dz d^2 \rho
\;
\Psi_{\lambda \bar \lambda}^V(z,\rho) 
\;
\sigma_{dip-A}(W,\rho)
\;
\Psi_{\lambda \bar \lambda}^{\gamma}(z,\rho)(z,\rho) \; .
\label{general_amplitude_gammaA_VA}
\end{equation}
In the equation above, $\lambda$ and $\bar \lambda$ are 
quark and antiquark helicities. Helicity conservation at high energy 
rescattering of the dipole in the nucleus is explicitly assumed.
The variable $\rho$ is the transverse size of the quark-antiquark dipole, 
and $z$ denotes the longitudinal momentum fraction carried by quark.
The longitudinal momentum fraction carried by antiquark is then $(1-z)$.
Using explicit formulae for photon and vector meson wave functions, 
the generic formula (\ref{general_amplitude_gammaA_VA}) can be written 
in a convenient way (for calculation see \cite{GM2006}).
The dipole-nucleus cross section can then be expressed in the
Glauber-Gribov picture in terms of the nuclear thickness
$T_A(b_{\gamma})$, as seen be the $q \bar q$ dipole in its way 
through the nucleus,
and the dipole-proton $\sigma_{dip-p}(\rho)$ cross section as
(see e.g., \cite{NZ}):
\begin{equation}
\sigma_{dip-A}(\rho,W) = 2 \int d^2 b_{\gamma}
\left\{
1 - \exp \left( -\frac{1}{2} T_A(b_{\gamma}) \sigma_{dip-p}(\rho,W) \right)
\right\} \; .
\label{dipole_nucleus_cross_section}
\end{equation}
This simple formula allows for an easy and convenient way to include
rather complex multiple scattering of the quark-antiquark dipole 
in the nucleus.
Several parametrizations of the dipole-nucleon cross section
have been proposed in the literature. 
%Here we shall consider a few simple parametrizations of 
%the dipole-nucleon cross section in order 
%to quantify the related theoretical uncertainties.
Most of them were obtained through fitting HERA deep-inelastic
scattering data which, in principle, does not allow for unique
extraction of the functional form. The saturation inspired
parametrizations are the most popular and topical at present.

\begin{figure}[!h]
\includegraphics[scale=0.4]{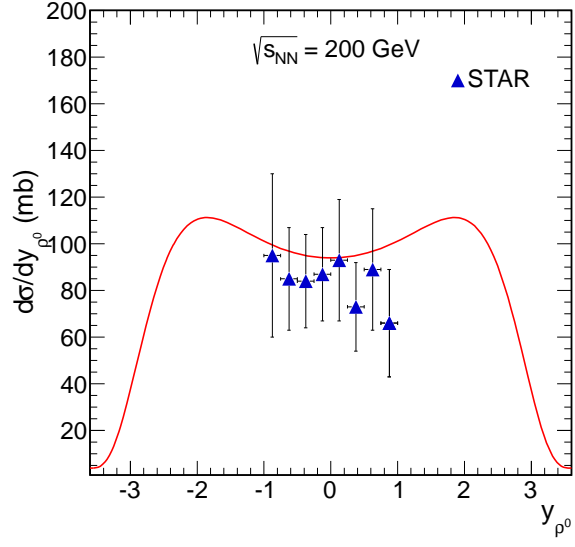}
   \caption{
\small Distribution in $\rho^0$ rapidity for single $\rho^0$ production.
       The STAR experimental data are taken from Ref. \cite{STAR2008_rho0}.}
 \label{fig:dsigma_V_dy}
\end{figure}

Before we go to double $\rho^0$ production, we shall briefly
show as an example the results for single $\rho^0$ production.
In Fig. \ref{fig:dsigma_V_dy} we present distributions in rapidity.
We obtain similar results as in other calculations in the literature
\cite{KN1999,FSZ2003,GM2005}.
Given the approximate character of the model
(see also Eq. (\ref{gammaA_rho0A})) the agreement is rather satisfactory.
Our total cross section equals 596 mb,
compared to 590 mb in the original Klein-Nystrand model \cite{KN1999}.  
The results of models presented in \cite{FSZ2003} and \cite{GM2005} exceed 
the STAR experimental data \cite{STAR2008_rho0}.
%-----------------------------------------------------------------------------
%-----------------------------------------------------------------------------

%-----------------------------------------------------
\subsection{Double scattering mechanism}
%-----------------------------------------------------

\begin{figure}[!h]
\includegraphics[scale=0.4]{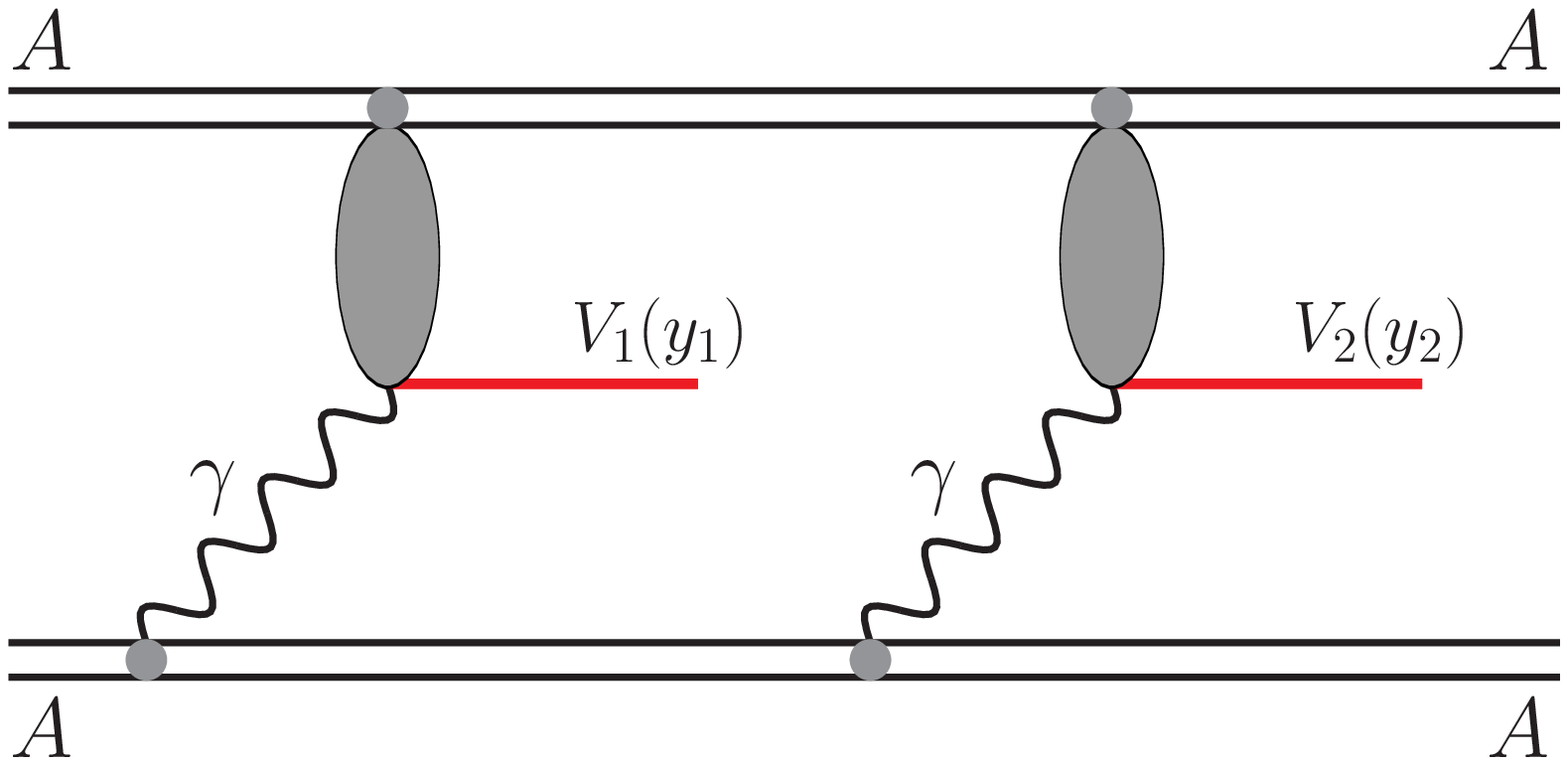}
\includegraphics[scale=0.4]{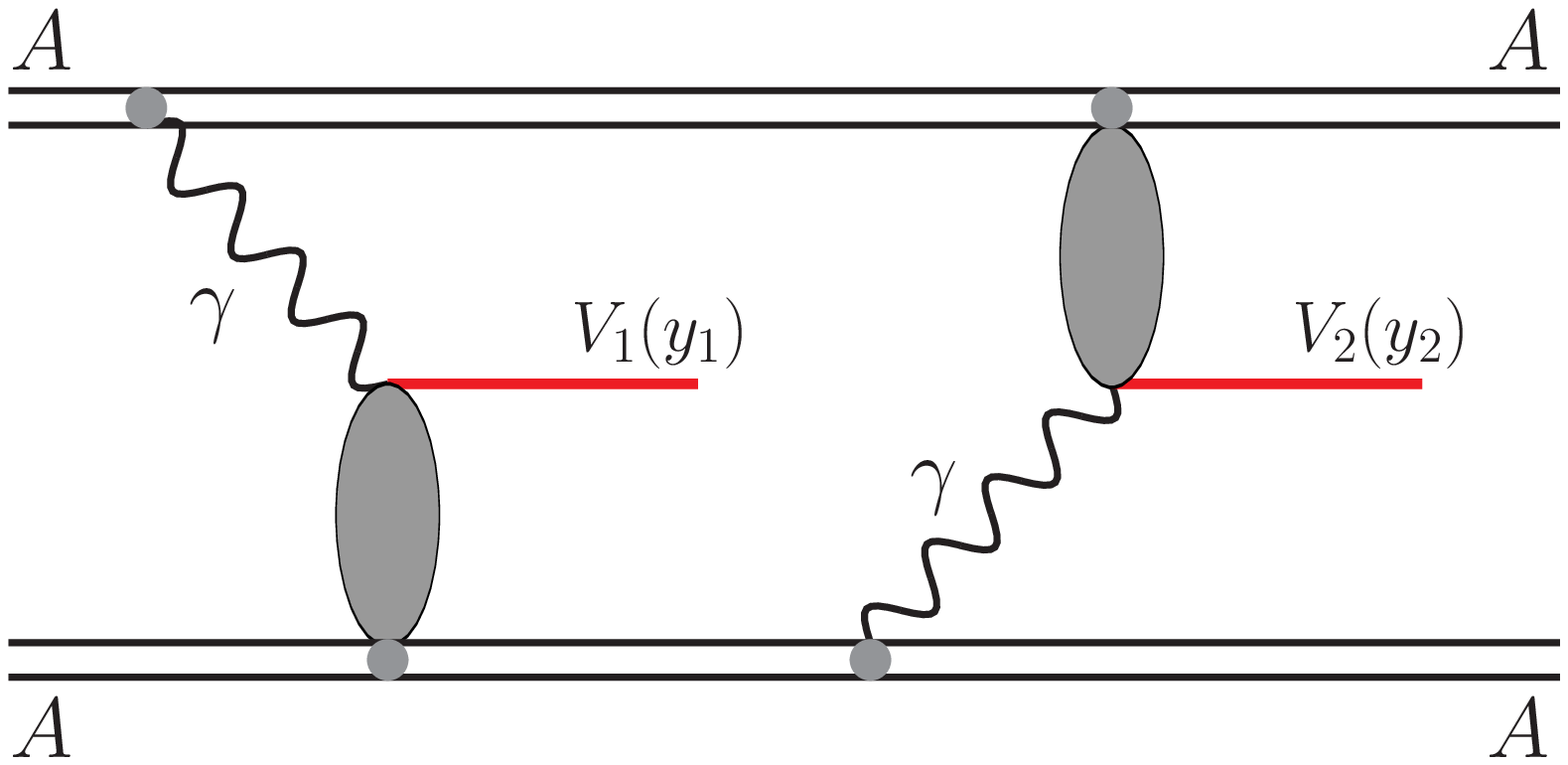}
\\
\includegraphics[scale=0.4]{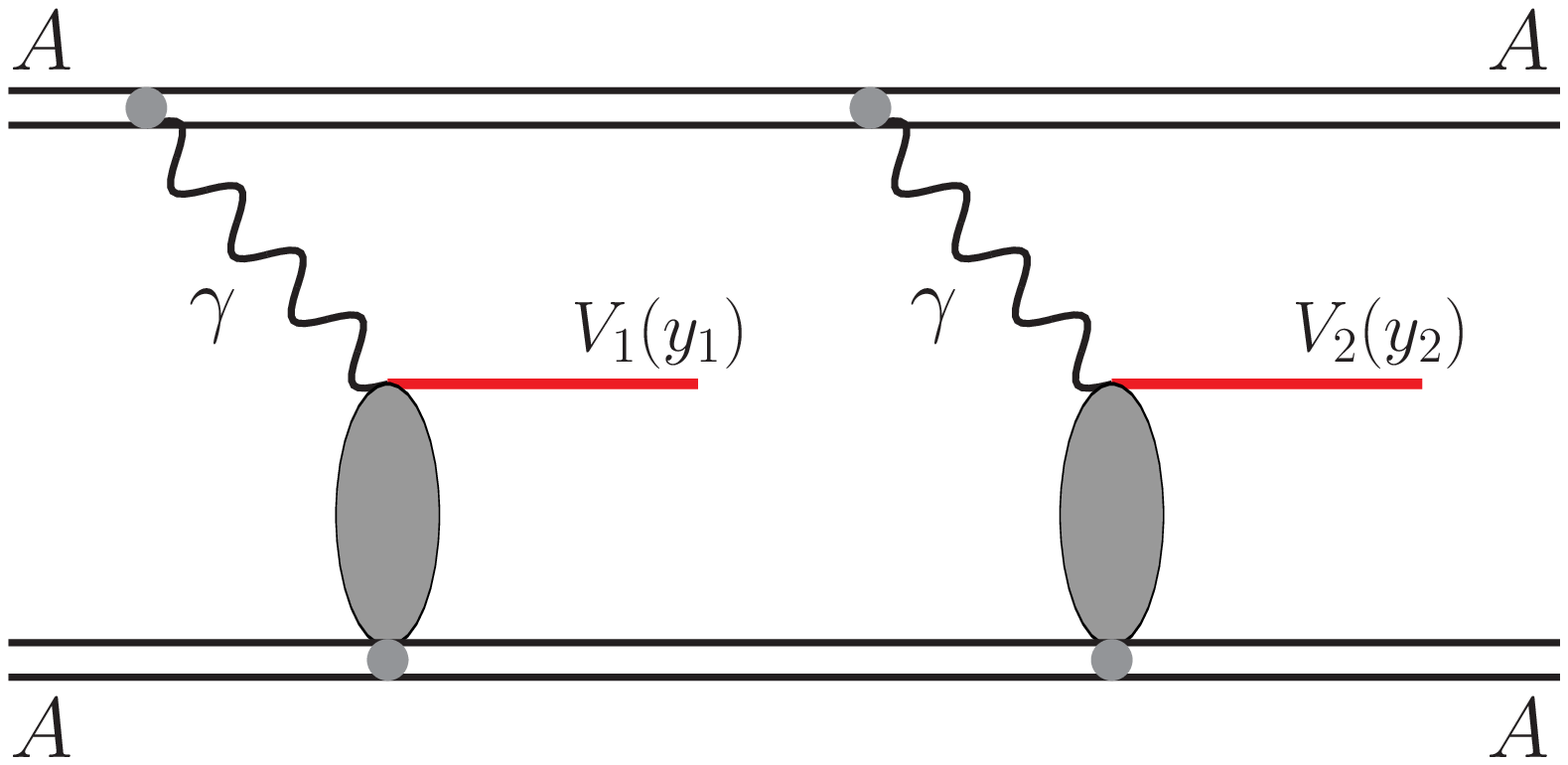}
\includegraphics[scale=0.4]{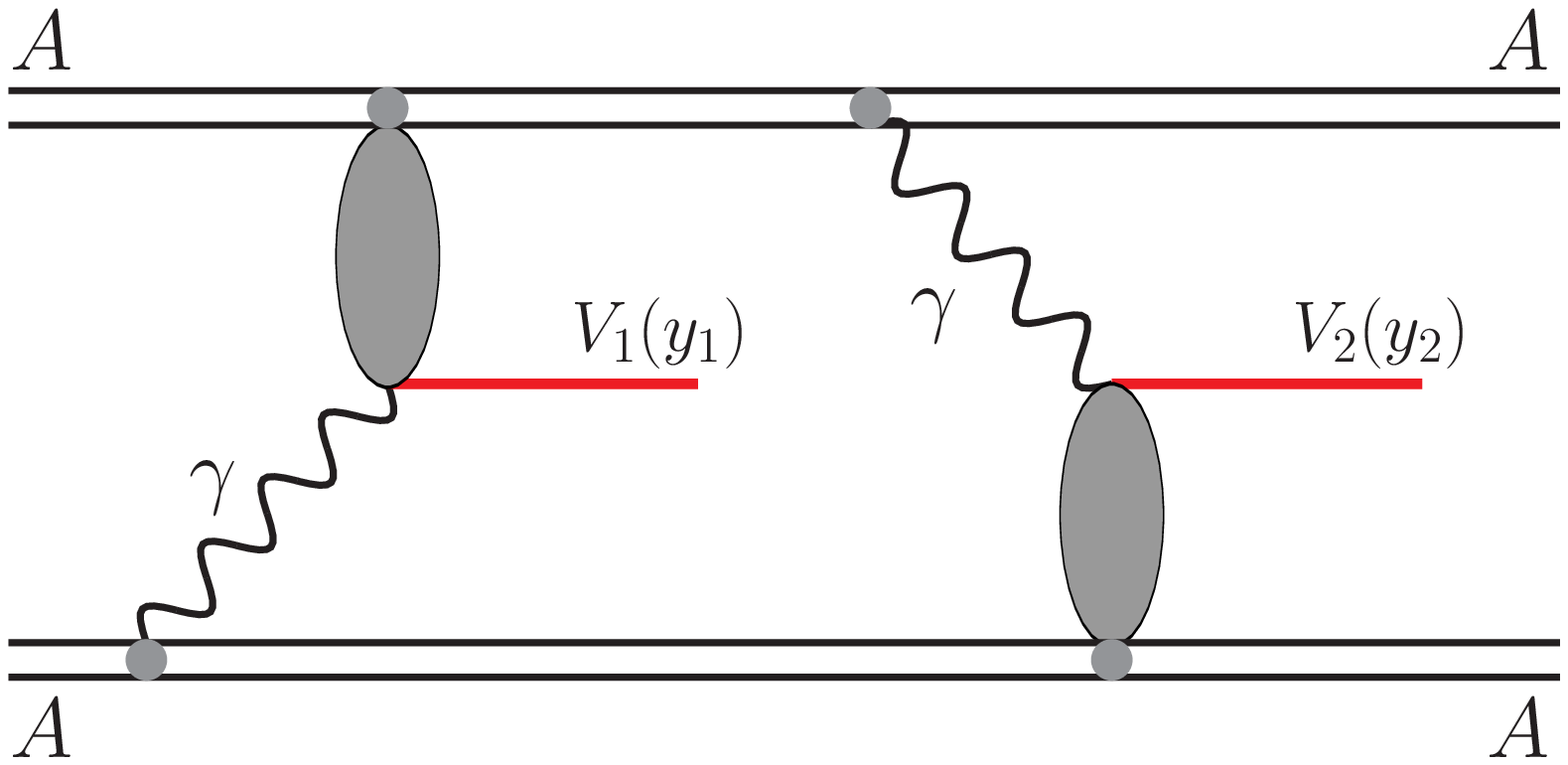}
   \caption{
\small The double-scattering mechanisms of two vector meson
production in ultrarelativistic, ultraperipheral collisions.
The blobs denote multiple scattering of quark-antiquark dipoles or
hadronic meson-like photon in the nucleus termed here ``pomeron exchange''
for brevity.
}
\label{fig:double_scattering}
\end{figure}

The generic diagrams of double-scattering production via photon-pomeron
or pomeron-photon exchange \footnote{By ``pomeron exchange'' we mean here
rather high-energy multiple diffractive rescattering of quark-antiquark
pairs (see e.g., \cite{CSS2012}) or virtual vector mesons.} 
mechanism are shown in Fig. \ref{fig:double_scattering}.

%-----------------------------------------------------------------------------
%-----------------------------------------------------------------------------

The double scattering process was discussed only in Ref.~\cite{KN1999},
where only a probabilistic formula for double and multiple vector meson
production was given. For example the cross section for double
scattering can be written as:
\begin{equation}
\sigma_{A A \to A A V_1 V_2}(\sqrt{s_{NN}}) = 
C \int S_{el}^2(b) P_{V_1}(b,\sqrt{s_{NN}}) P_{V_2}(b,\sqrt{s_{NN}}) d^2 b \; .
\label{sigma_AA_AAV}
\end{equation}
In the equation above $b$ is the impact parameter (transverse distance 
between nuclei). We have included natural limitations in the impact
parameter
\begin{equation}
S_{el}^2(b) = \exp \left( -\sigma_{NN}^{tot} T_{A_1 A_2}(b) \right)
\approx \theta\left( b - (R_1+R_2) \right)  \;  .
\end{equation}
It may be interpreted as a survival probability for nuclei not to
break up.
The probability density of single meson production is
\begin{equation}
P_V(b,\sqrt{s_{NN}}) = \frac{d \sigma_{AA \to AAV}(b;\sqrt{s_{NN}})}
{2 \pi b db} \; .
\label{probability}
\end{equation}
The constant $C$ is in the most general case 1 or $\frac{1}{2}$ for 
identical vector mesons $V_1 = V_2$.
We have explicitly indicated the dependence of the probabilities on 
nucleon-nucleon energy.
The probability densities $P_V$ increase with increasing cm energy.

The photon flux factor is calculated here as:
\begin{equation}
\frac{d^3N}{d^2b d\omega} 
= \frac{Z^2\alpha_{em}X^2}{\pi^2\omega b^2} K_1^2(X)  \; ,
\end{equation}
where 
$X = \frac{b \omega}{\gamma}$.
We leave a more refined treatment for future studies.
 
The simple formula (\ref{sigma_AA_AAV}) can be generalized to calculate 
two-dimensional distributions in rapidities of both vector mesons
\begin{eqnarray}
\frac{d \sigma_{AA \to AA V_1 V_2}}{d y_1 d y_2} = C \int 
 && \left( \frac{d P_1^{\gamma \Pom}(b,y_1;\sqrt{s_{NN}})}{d y_1}
               + \frac{d P_1^{\Pom \gamma}(b,y_1;\sqrt{s_{NN}})}{d y_1}
               \right)
\nonumber \\
\times 
 && \left( \frac{d P_2^{\gamma \Pom}(b,y_2;\sqrt{s_{NN}})}{d y_2}
                + \frac{d P_2^{\Pom \gamma}(b,y_2;\sqrt{s_{NN}})}{d y_2}
               \right)   \;  d^2 b  \; . 
\label{dsigma_dy1dy2}    
\end{eqnarray}
$P_1$ and $P_2$ are probability densities for producing
one vector meson $V_1$ at rapidity $y_1$ and the second vector meson $V_2$
at rapidity $y_2$ for fixed impact parameter $b$ 
of the heavy-ion collision. 
Then the differential probability density can be written as: 
\begin{equation}
\frac{d P_V(b,\sqrt{s_{NN}})}{dy} =
 \frac{d \sigma_{AA \to AAV}(b;\sqrt{s_{NN}})}
{2 \pi b db dy} \; .
\label{generalized_probability}
\end{equation}
The produced vector mesons in each step are produced in very 
broad range of (pseudo)rapidity \cite{KN1999,GM2006} and extremely
small transverse momenta.

%---------------------------------------------------
\subsection{Smearing the $\rho^0$ masses}
%---------------------------------------------------

The $\rho^0$ resonance is fairly broad.
We consider two different approximations: (a) fixed $\rho^0$ mass,
(b) smeared mass.
In the fixed-mass approximation, the mass of the di-meson system 
(for identical vector meson masses) can be calculated from 
the simple formula
\begin{equation}
M_{\rho^0\rho^0}^2 = 2 m_{\rho^0}^2 \left(1+\cosh(y_1-y_2)\right) \; .
\end{equation}
The mass is then calculated for each phase space point ($y_1, y_2$) 
(see Eq. (\ref{dsigma_dy1dy2})) and put into a histogram.

In a more refined approximation (b) one has to include in addition a
smearing of the $\rho^0$ mass. Then the cross section can be written as:
\begin{equation}
\frac{d \sigma_{AA \to AA \rho_0^* \rho_0^*}}{dm_1 dm_2 dy_1 dy_2}
= f(m_1) f(m_2) 
\frac{d \sigma_{AA \to AA \rho_0^* \rho_0^*}}{dy_1 dy_2}
(y_1 y_2; m_1, m_2)  \; ,
\end{equation}
where $m_1$ and $m_2$ are the running masses of $\rho^0$ mesons and
$f_1(m_1)$ and $f_2(m_2)$ are smearing distributions spectral shapes as used below.
The last term is the production cross section for running $\rho^0$ meson masses $m_1$ and $m_2$.
In general, the cross section depends on the values of the $\rho^0$ masses.
The smaller the mass, the larger the cross section.
In the present analysis the spectral shapes are calculated as:
\begin{equation}
f(m) = |{\cal A}|^2 / \int |{\cal A}|^2 dm \; ,
\label{spectral_shape}
\end{equation}
where the amplitude is, as often done in the literature, parametrized 
in the form:
\begin{equation}
{\cal A} = {\cal A}_{BW} \frac{\sqrt{m m_{\rho} \Gamma(m)}}
{m^2 - m_{\rho}^2 + i m_{\rho} \Gamma(m)} + {\cal A}_{\pi \pi}  \; .
\end{equation}
The mass-dependent width is parametrized as in \cite{STAR2008_rho0}
\begin{equation}
\Gamma(m) = \Gamma_{\rho} \frac{m_{\rho}}{m}
\left( 
\frac{m^2 - 4 m_{\rho}^2}{m_{\rho}^2 - 4 m_{\pi}^2}
\right)^{3/2} \; ,
\label{width}
\end{equation}
i.e. vanishes below the two-pion threshold and as a consequence
also the spectral shape vanishes below the two-pion threshold.
The extra constant is often added to describe a big asymmetry (enhancement
of the left hand side of the $\rho^0$ resonance). A physical interpretation
of the constant term for proton-proton collisions
can be found in Ref.~\cite{SS2005} where it was described as due to
the Deck two-pion continuum.

In the following we shall parametrize the first term 
in Eq. (\ref{gammaA_VA}) for the resonance mass ($m_{\rho}$ = 770 MeV) 
(see \cite{KN1999}).
However, we shall include the running masses ($m_1$ and $m_2$) in 
calculating $t_{max}$.
This leads to a deformation of the spectral shapes of resonances.
One could also include the dependence of the production cross section
on running masses which would lead to an extra deformation. These
deformations have an influence on the actual value of the integrated 
cross section, which will be discussed in the next section.
%-----------------------------------------------------------------------------
\begin{figure}[!h]
\includegraphics[scale=0.4]{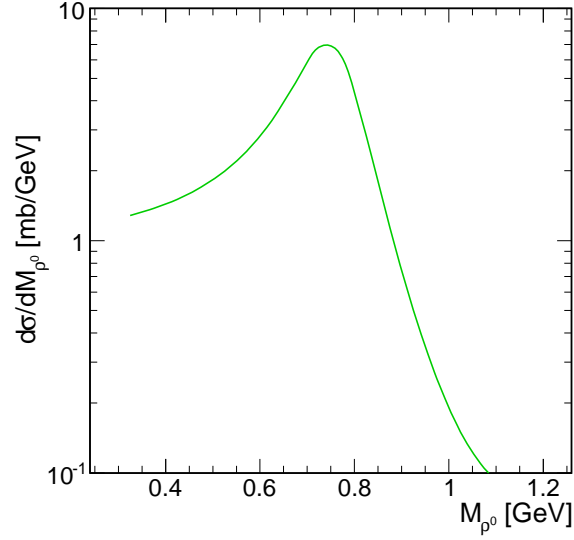}
   \caption{
\small Distribution in the mass of one of $\rho^0$ meson
produced in double scattering mechanism. An enhancement at
lower $M_{\rho^0}$ is clearly visible.}
 \label{fig:dsig_dmrho}
\end{figure}
%-----------------------------------------------------------------------------

In Fig.~\ref{fig:dsig_dmrho} we show the invariant-mass distribution 
of single $\rho^0$ production. The $\rho^0$ peak is
asymmetric. Clearly the left hand flank is enhanced due to production
mechanism discussed in the previous section.
This asymmetry is opposite than in many other processes.

%-----------------------------------------------------------------------------
\begin{figure}[!h]
\includegraphics[scale=0.4]{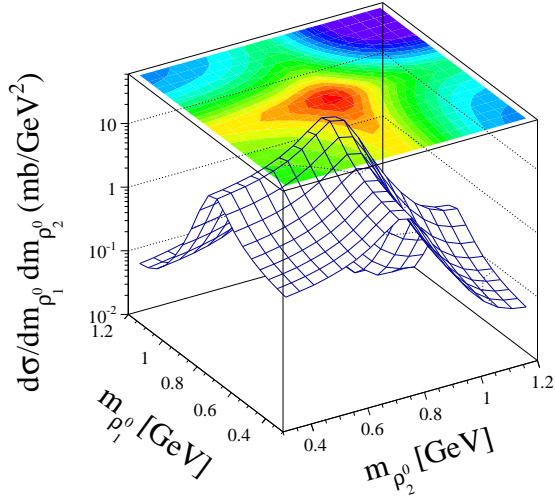}
   \caption{
\small Two-dimensional distribution in $\rho^0$ meson masses
produced in the double scattering mechanism. An enhancement at
lower $m_1$ and $m_2$ is clearly visible.}
 \label{fig:dsig_dmrho1dmrho2}
\end{figure}
%-----------------------------------------------------------------------------

In Fig.~\ref{fig:dsig_dmrho1dmrho2} we show two-dimensional distributions
in masses of both $\rho^0$ mesons for double-$\rho^0$ production. 
A strong enhancement at low masses is
again clearly visible. The enhancement of the low masses may lead
to an enhancement of the cross section compared to fixed-mass
calculation. This will be discussed in the section \ref{results} 
where our results will be presented.

%-------------------------------------------------
\subsection{$\rho^0 \to \pi^+ \pi^-$ decays}
%-------------------------------------------------

The $\rho^0$ mesons from photoproduction are dominantly transversely 
polarized and have negligibly small transverse momenta with respect
to the direction of heavy ions. We parametrize the decay function 
in the $\rho^0$ center of mass as:
\begin{equation}
f(\theta^*) = \frac{3}{2} sin^2(\theta^*) \; .
\label{decay_function}
\end{equation}
The calculations are done as follows. First a map of the cross section
for a dense grid in ($y_1, y_2$) is prepared in the case of double
scattering with fixed $\rho^0$ meson masses (the corresponding $\rho^0$
mesons have negligibly small transverse momenta) or in 
($y_1, y_2, m_1, m_2$) when the smearing of the $\rho^0$ mass is taken
into account.
For the $\gamma \gamma$ mechanism one has to take into account 
in adition the transverse momenta of the $\rho^0$ mesons as dictated 
by the VDM-Regge or two-gluon exchange models.

Then the decays are done in a separate Monte Carlo code. The
distributions in $\rho^0$ centers of mass are generated randomly with 
the decay function given by (\ref{decay_function}) or isotropically. 
Next a Lorentz transformations to the overall ion-ion center of mass 
(laboratory system for both RHIC and LHC) is performed. 
Different kinematical variables related to charged pions are calculated 
and corresponding distributions are obtained by an appropriate binning. 
Since we have the full kinematics of the event any cut on kinematical variables 
can be easily imposed.

%--------------------------------------------
\section{First numerical results}
%--------------------------------------------
\label{results}

Having fixed the details of single vector meson production
we can now proceed to the production of two vector mesons.
For the RHIC energy we consider $^{197}Au + ^{197}Au$ collisions and 
for the LHC energy we take into account $^{208}Pb + ^{208}Pb$ collisions.
As an example in Table~\ref{tab:sig_rhorho} we show total cross section
for $\rho^0 \rho^0$ production in ultraperipheral, ultrarelativistic
heavy ion collisions. The double scattering cross section 
for $\rho^0\rho^0$ pair production at RHIC energy
is about 1.5 mb. This is a rather large cross section (compared to the 
cross section for exclusive production of $\rho^0 \rho^0$
via photon-photon fusion which at RHIC energy $\sqrt{s_{NN}}$ = 200 GeV
is of the order of 0.1 mb)\footnote{For comparision the cross section 
for $Pb Pb \to Pb Pb \pi^+ \pi^-$
via photon-photon fusion at $\sqrt{s_{NN}}$ = 3.5 TeV is 46.7 mb \cite{KS2013}.}.
More details can be found in our previous paper \cite{KSS2009} and 
in Table~\ref{tab:sig_rhorho}. The cross section for the smeared
$\rho^0$ masses is larger than that for larger than that for fixed
resonance masses. The cross section for the VDM-Regge contribution
is rather small. Its relative contribution should increase
at LHC energies where the photon-photon luminosities are much larger.

%===========================================================================================
\begin{table}[!h]
\caption{Cross sections (in mb) for single $\rho^0$ production
and double scattering and photon-photon mechanisms 
of $\rho^0\rho^0$ production for fixed and smeared mass of $\rho^0$ meson.}
\begin{center}
\begin{tabular}{||l| c| c||}
\hline \hline
Energy								&    $m_{\rho^0}$ = 0.77549 GeV	& Mass smearing	\\ \hline

RHIC ($\sqrt{s_{NN}}$ = 200 GeV), single $\rho^0$ production		&   	596   			&		\\ 
 LHC ($\sqrt{s_{NN}}$ = 3.5 TeV), single $\rho^0$ production		&       4000  			&    	   	\\ 
 LHC ($\sqrt{s_{NN}}$ = 5.5 TeV), single $\rho^0$ production		&       4795  			&    	   	\\ \hline

RHIC ($\sqrt{s_{NN}}$ = 200 GeV), double scattering		&   	1.5   			&	1.55	\\ 
 LHC ($\sqrt{s_{NN}}$ = 3.5 TeV), double scattering 		&         			&    	15.25   \\ 

RHIC ($\sqrt{s_{NN}}$ = 200 GeV), double scattering, $|\eta_\pi|<1$&   	   			&	0.15	\\ 
 LHC ($\sqrt{s_{NN}}$ = 3.5 TeV), double scattering, $|\eta_\pi|<1$&         			&    	0.3   \\ \hline

RHIC ($\sqrt{s_{NN}}$ = 200 GeV), $\gamma \gamma$, VDM-Regge   	& 	7.5 10$^{-3}$ 		&      		\\ 
RHIC ($\sqrt{s_{NN}}$ = 200 GeV), $\gamma\gamma$, low-energy bump & 	95 10$^{-3}$ 		&       	\\  
RHIC ($\sqrt{s_{NN}}$ = 200 GeV), $\gamma \gamma$, VDM-Regge, $|\eta_\pi|<1$   	& 	0.5 10$^{-3}$ 		&      		\\ 
RHIC ($\sqrt{s_{NN}}$ = 200 GeV), $\gamma\gamma$, low-energy bump, $|\eta_\pi|<1$ & 	14.6 10$^{-3}$ 		&       	\\ 
\hline \hline

\end{tabular}
\end{center}
\label{tab:sig_rhorho}
\end{table}
%============================================================================================
%
%-----------------------------------------------------------------------------
\begin{figure}[!h]
\includegraphics[scale=0.4]{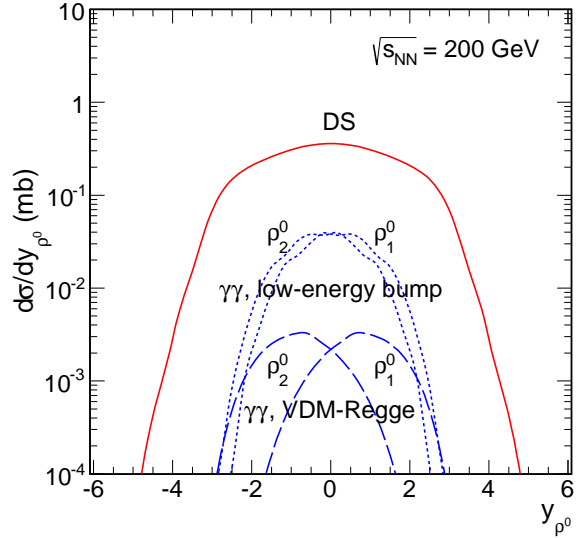}
   \caption{
\small Rapidity distribution of one of $\rho^0$ mesons
produced in double scattering mechanism. The double-scattering
contribution is shown by the solid (red online) line and the dashed
lines (blue online) represents distributions of forward and backward 
$\rho^0$ produced in the VDM-Regge photon-photon fusion.}
 \label{fig:dsig_dyrho}
\end{figure}
%-----------------------------------------------------------------------------

Distributions in $\rho^0$ meson rapidity is shown 
in Fig.~\ref{fig:dsig_dyrho}. One can observe a clear dominance
of the double scattering component over the photon-photon component. 
At the LHC the proportions should be slightly different.
For the photon-photon mechanism we show separate contributions 
for the forward and backward $\rho^0$ mesons.

%-----------------------------------------------------------------------------
\begin{figure}[h]
 \begin{minipage}[t]{0.45\textwidth}
  \centering
  \includegraphics[width=1.\textwidth]{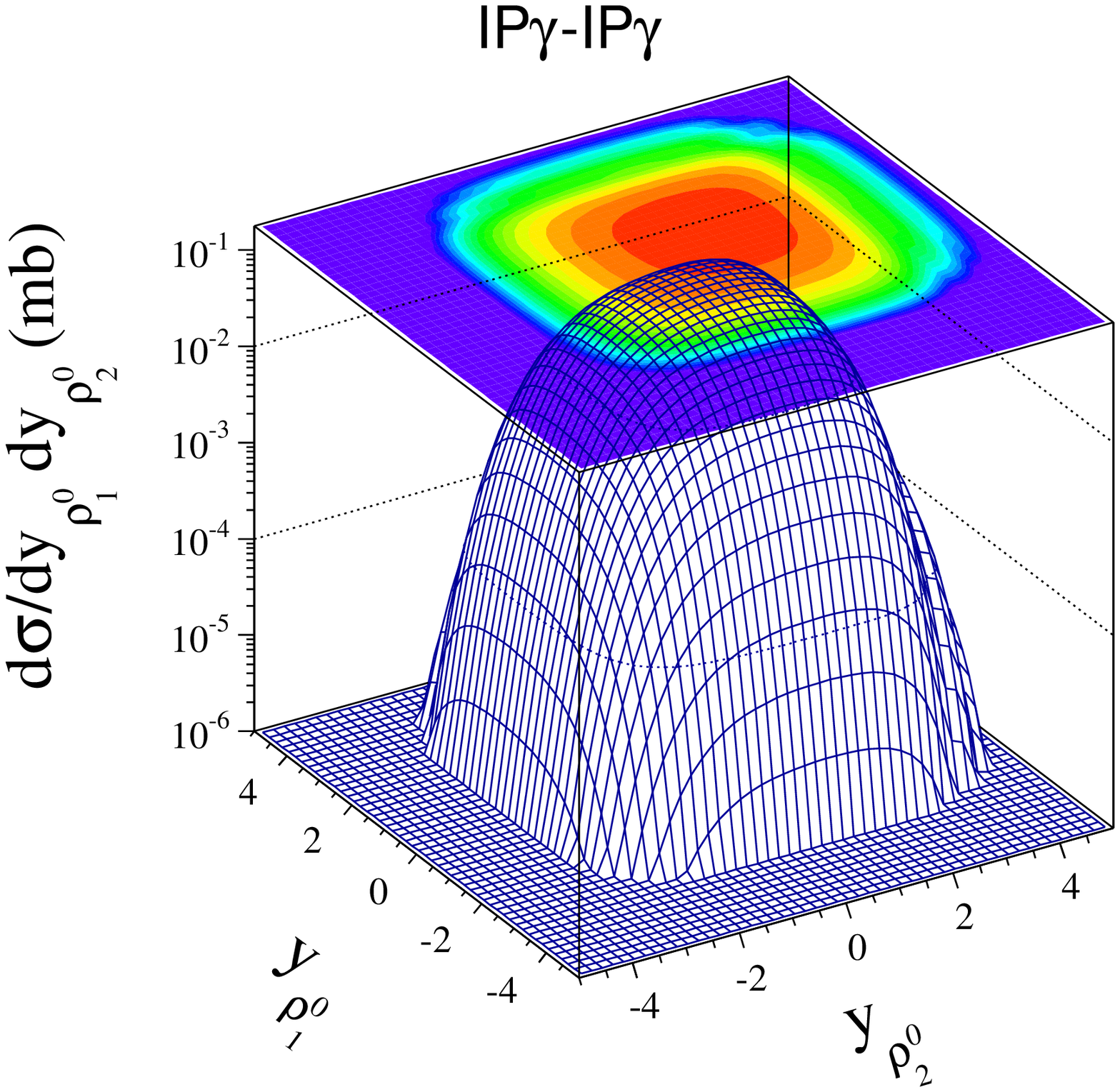}
 \end{minipage}
 \hspace{0.03\textwidth}
 \begin{minipage}[t]{0.45\textwidth}
  \centering
  \includegraphics[width=1.\textwidth]{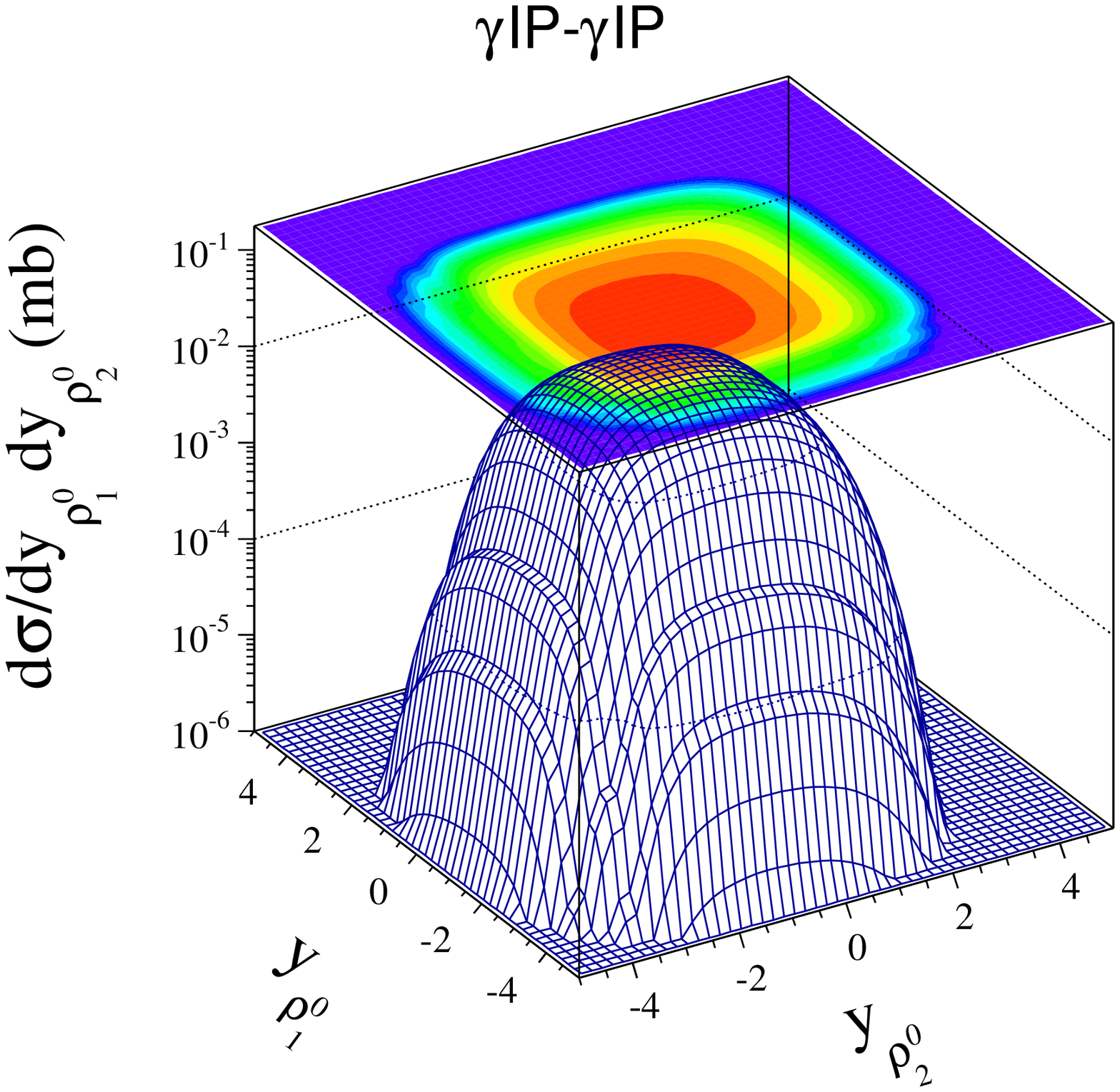}
 \end{minipage}
 \hspace{0.03\textwidth}
 \begin{minipage}[t]{0.45\textwidth}
  \centering
  \includegraphics[width=1.\textwidth]{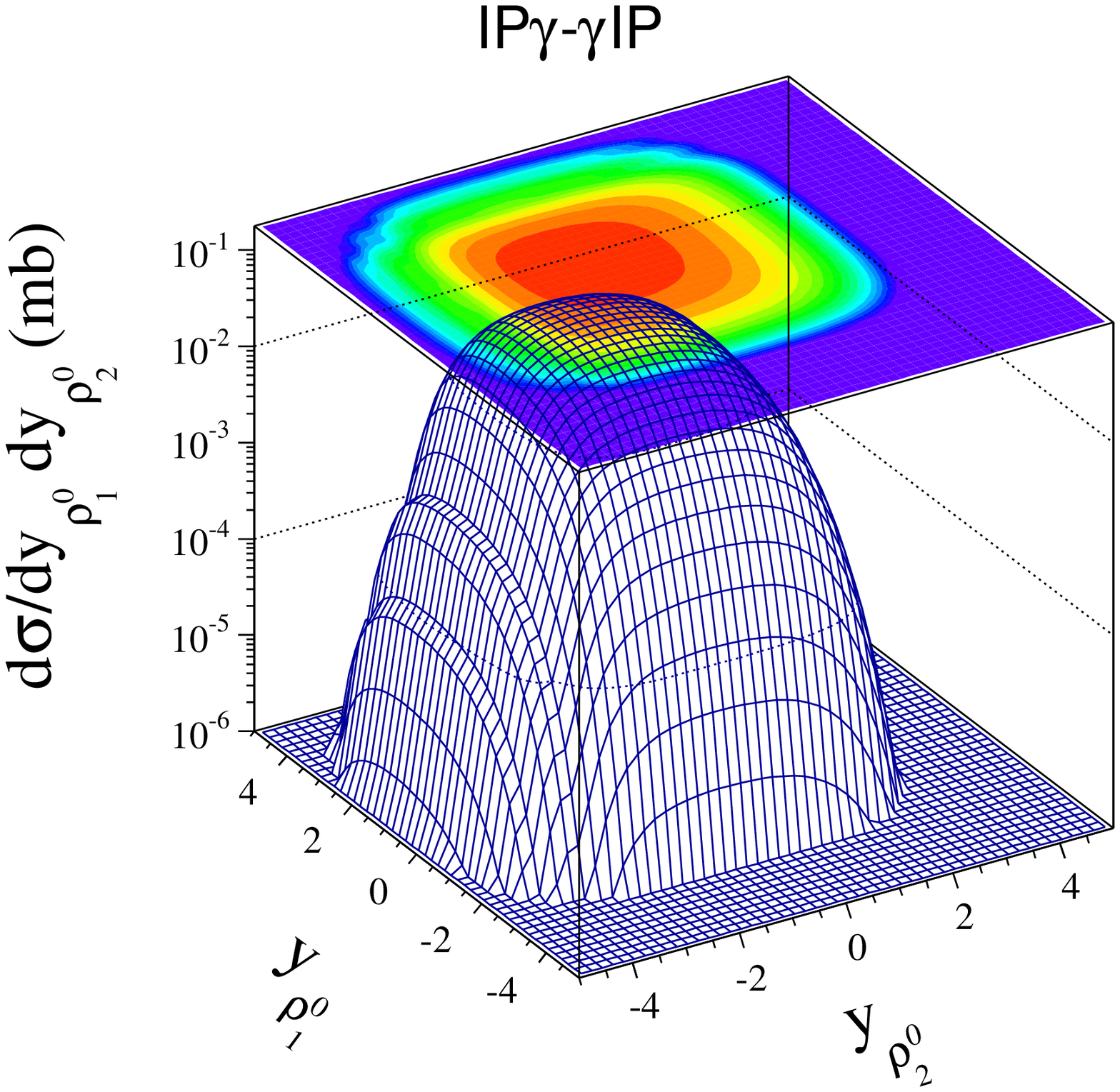}
 \end{minipage}
 \hspace{0.03\textwidth}
 \begin{minipage}[t]{0.45\textwidth}
  \centering
  \includegraphics[width=1.\textwidth]{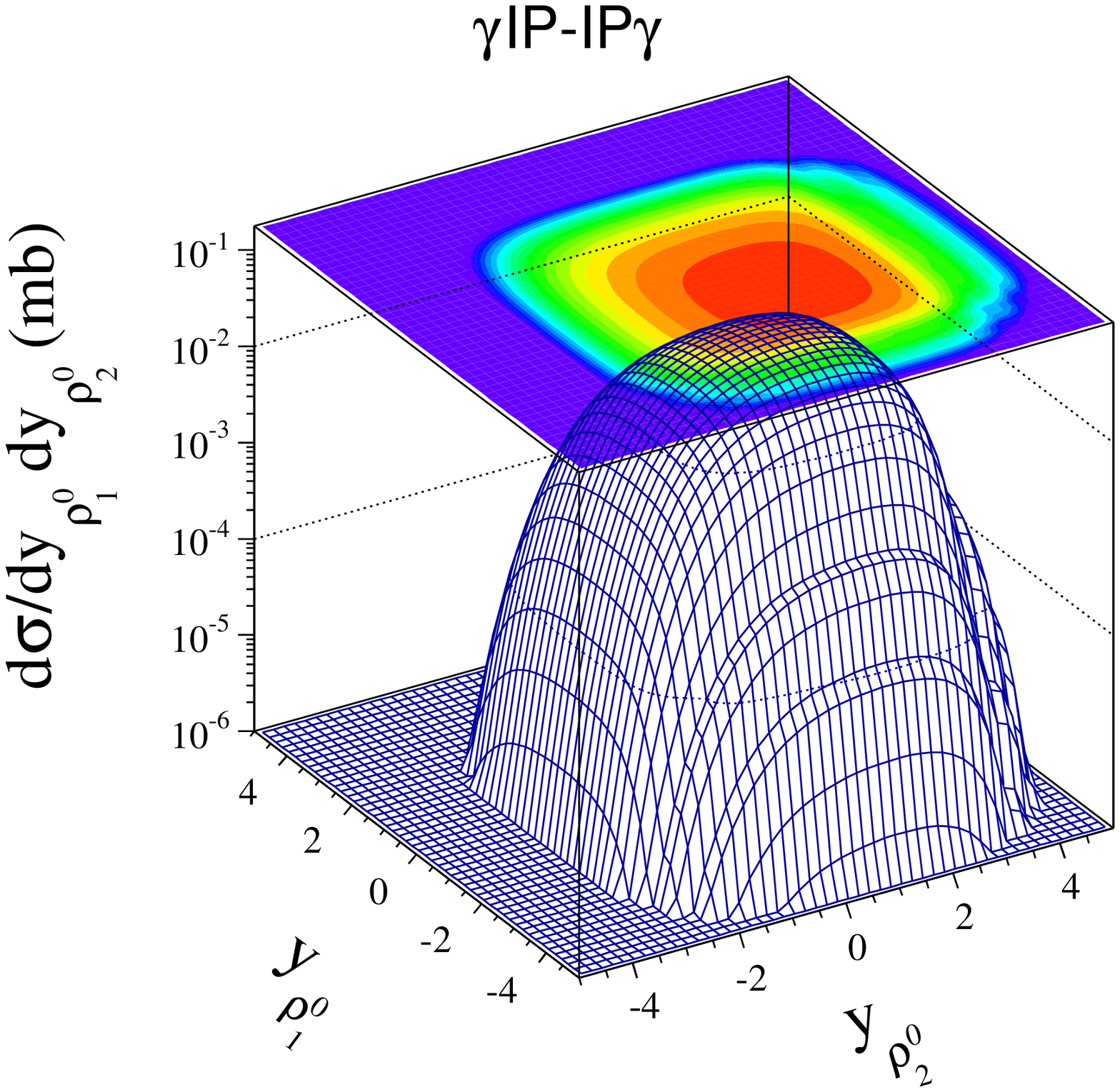}
 \end{minipage}
   \caption{
\small Contributions of individual diagrams of Fig. \ref{fig:double_scattering}
to two-dimensional distribution in $\rho^0$ meson rapidities
for double scattering production for the full phase space at 
$\sqrt{s_{NN}}$ = 200 GeV.}
 \label{fig:dsig_dy1dy2_4diagrams}
\end{figure}
%-----------------------------------------------------------------------------

Now we wish to discuss briefly the contributions of individual diagrams
of Fig.~\ref{fig:double_scattering} to $(y_1,y_2)$ two dimensional distribution.
The corresponding distributions are shown in Fig.~\ref{fig:dsig_dy1dy2_4diagrams}.
The distributions for different combinations are identical in shape but
located in different corners in $(y_1,y_2)$ space.

%-----------------------------------------------------------------------------
\begin{figure}[!h]
\includegraphics[scale=0.4]{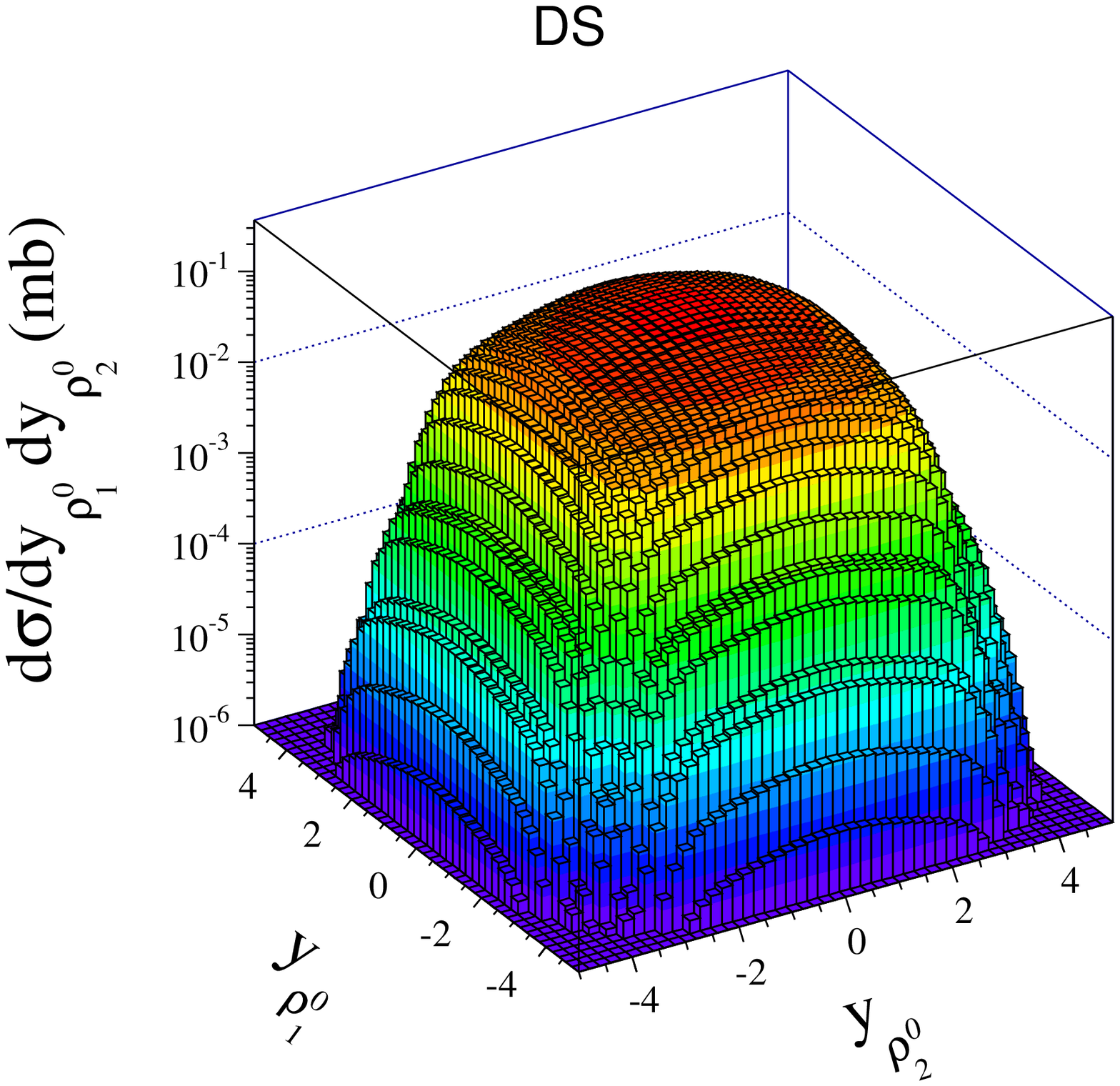}
\includegraphics[scale=0.4]{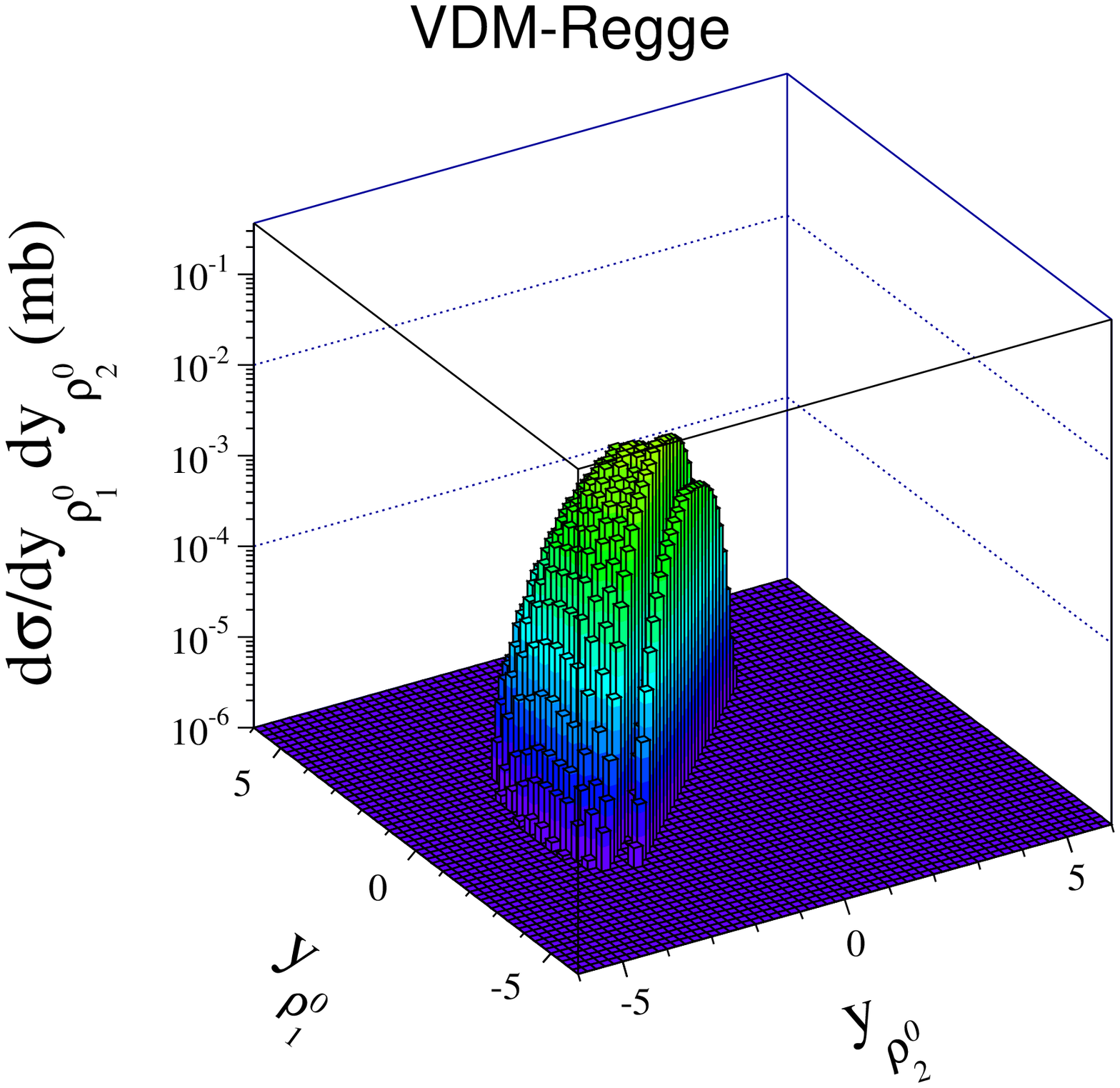}
   \caption{
\small Two-dimensional distribution $(y_1,y_2)$ in $\rho^0$ meson 
rapidities for double scattering (left panel) and VDM-Regge
photon-photon (right panel) production for the full phase space at 
$\sqrt{s_{NN}}$ = 200 GeV. The distribution for $\gamma\gamma$
subprocess is asymmetric because in this case the first $\rho^0$
is emitted in forward direction 
and the second $\rho^0$ is emitted in backward direction.}
 \label{fig:dsig_dy1dy2}
\end{figure}
%-----------------------------------------------------------------------------

A full (including all contributions) two-dimensional distribution 
in rapidity of each of the mesons is shown in Fig.~\ref{fig:dsig_dy1dy2} 
which in the approximations made is a sum of the individual
contributions. The distribution is rather
flat in the entire $(y_1, y_2)$ space (see the left panel). 
This is in contrast to the two-photon processes, where the cross section
is concentrated along the $y_1 = y_2$ diagonal (see the right panel).
In principle, this clear difference can be used to distinguish 
the double photoproduction from the photon-photon fusion.
The asymmetry with respect to $y_1=y_2$ line for the photon-photon
mechanism is due to our convention where $y_1$ denotes rapidity of 
the forward and $y_2$ rapidity of the backward emitted $\rho^0$ mesons.
This can be done only in model calculation.

%-----------------------------------------------------------------------------
\begin{figure}[!h]
\includegraphics[width=8cm]{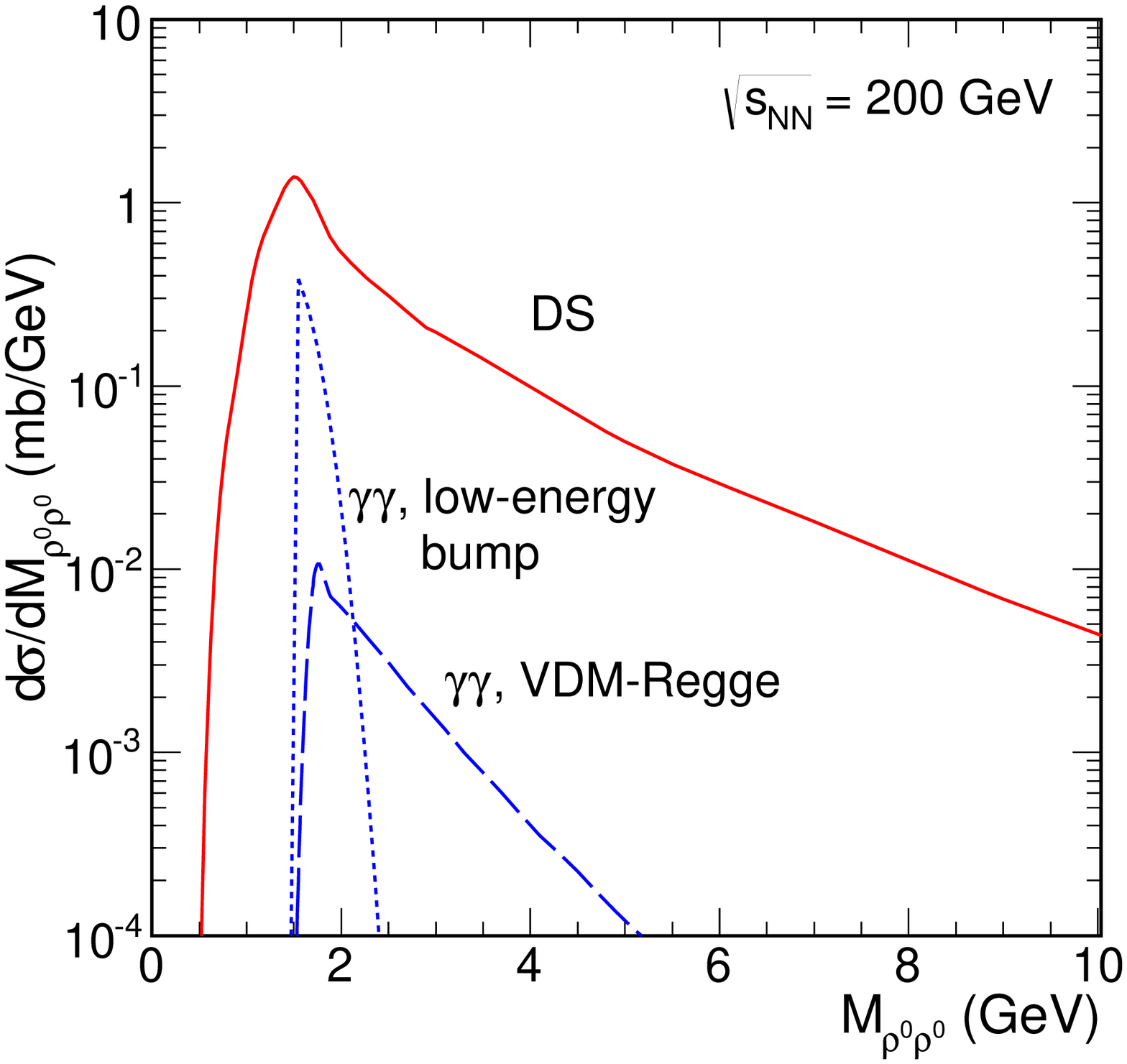}
   \caption{
\small Invariant mass distribution of $\rho^0 \rho^0$ for double
scattering (solid line), VDM-Regge photon-photon (dashed line)
and low-energy bump (dotted line) mechanisms
 for full phase space at $\sqrt{s_{NN}}$ = 200 GeV.}
 \label{fig:dsig_dMVV}
\end{figure}
%------------------------------------------------------------------------------

The corresponding distribution in the $\rho^0 \rho^0$ invariant mass is 
shown in Fig.~\ref{fig:dsig_dMVV}. In general, somewhat larger invariant masses
are generated via the double scattering mechanism than in two-photon processes. 
The reader is asked to compare the present plot with analogous plot in 
Ref.~\cite{KSS2009}.

%-----------------------------------------------------------------------------
\begin{figure}[!h]
\includegraphics[width=8cm]{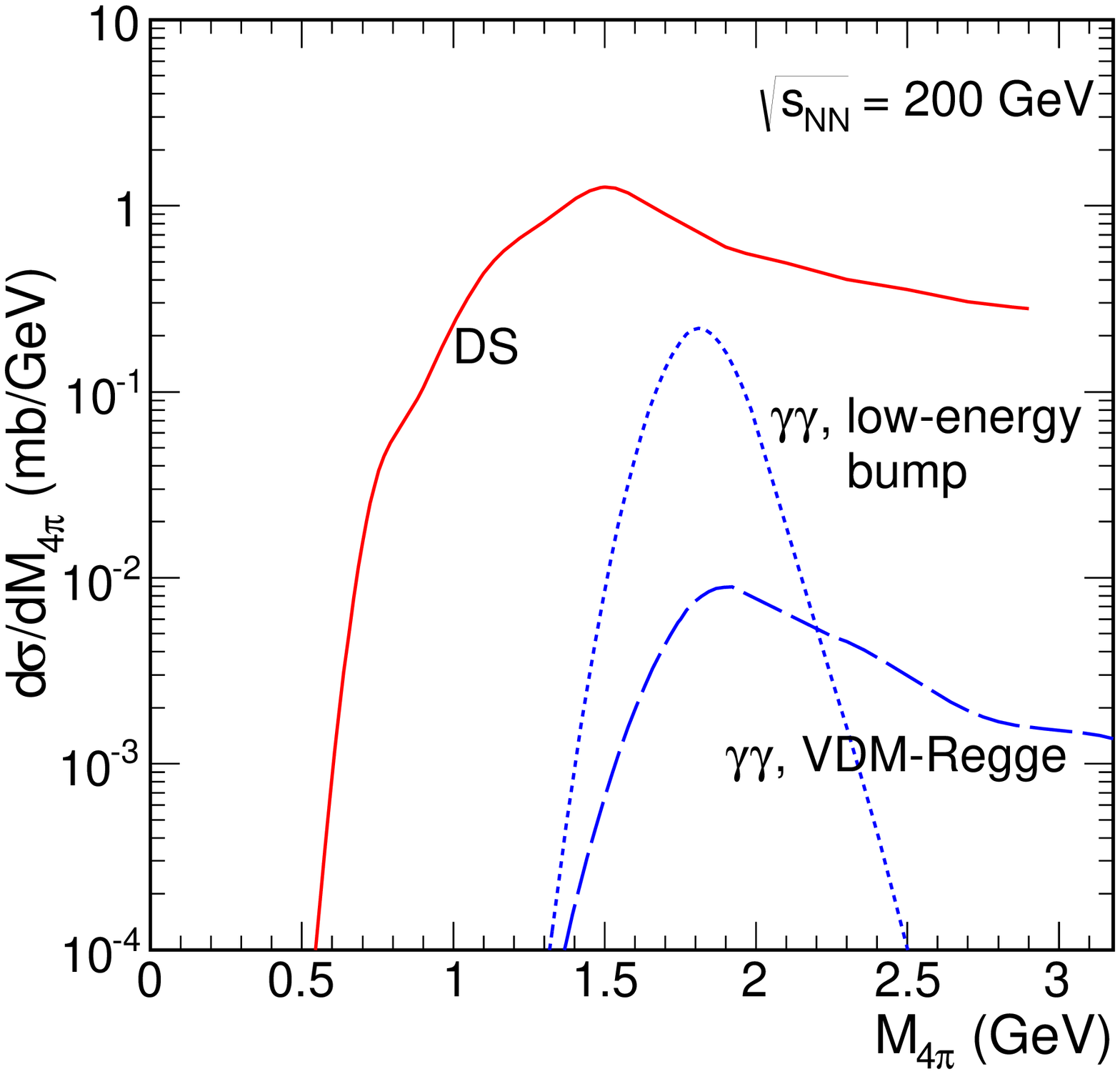}
\includegraphics[width=8cm]{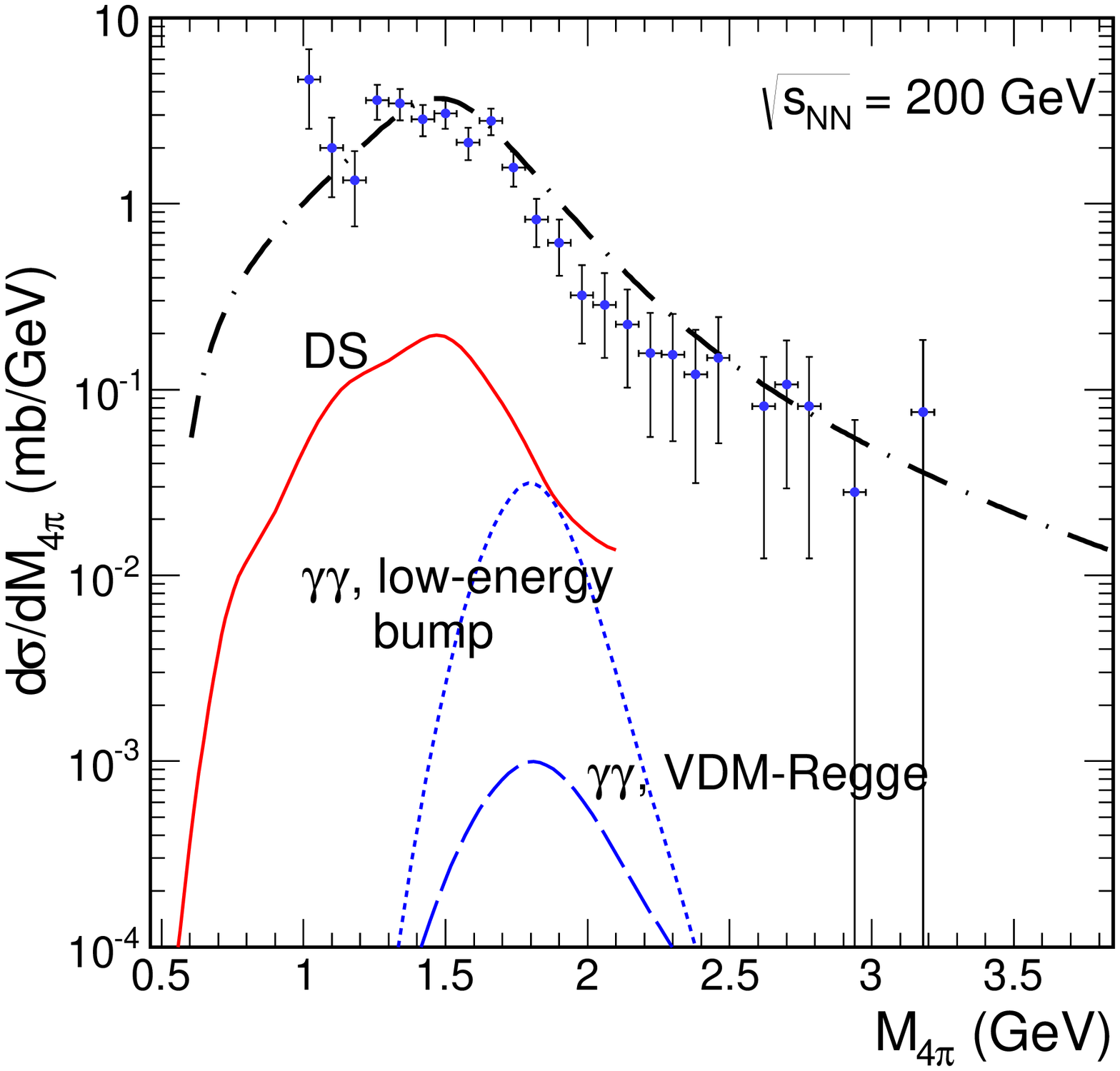}
   \caption{
\small Four-pion invariant mass distribution for double
scattering mechanism (solid line), VDM-Regge photon-photon (dashed line)
and low-energy bump (dotted line) mechanisms 
for full phase space (left panel) and for the limited acceptance 
STAR experiment (right panel) at $\sqrt{s_{NN}}$ = 200 GeV. 
The STAR experimental data \cite{STAR2010_4pi} have been corrected
by acceptance function \cite{BGpc}. The dash-dotted line represents
a fit of the STAR collaboration.}
 \label{fig:dsig_dM4pi}
\end{figure}
%------------------------------------------------------------------------------

In real experiments, charged pions are measured rather than $\rho^0$ mesons.
Therefore, we now proceed to a presentation of some observables related 
to charged pions. We start from the presentation of four-pion invariant mass
distribution (see Fig.~\ref{fig:dsig_dM4pi}). The distribution for the
whole phase space extends to large invariant masses, while the
distribution in the limited range of (pseudo)rapidity spanned by the STAR
detector give a shape similar to the measured distribution (see
dash-dotted line in the right panel of Fig. \ref{fig:dsig_dM4pi}).
However, the double-scattering contribution accounts only for 20 \% of 
the cross section measured by the STAR collaboration \cite{STAR2010_4pi}. 
Apparently, the production of the $\rho^0(1700)$ resonance and its 
subsequent decay into the four-pion final state (see e.g., \cite{PDG}) 
is the dominant effect for the limited STAR acceptance. Both, the production 
mechanism of $\rho^0(1700)$, and its decay into four charged pions are 
not yet understood. We therefore leave the modeling of these production 
and decay processes for a separate study.

%-----------------------------------------------------------------------------
\begin{figure}[!h]
\includegraphics[width=8.0cm]{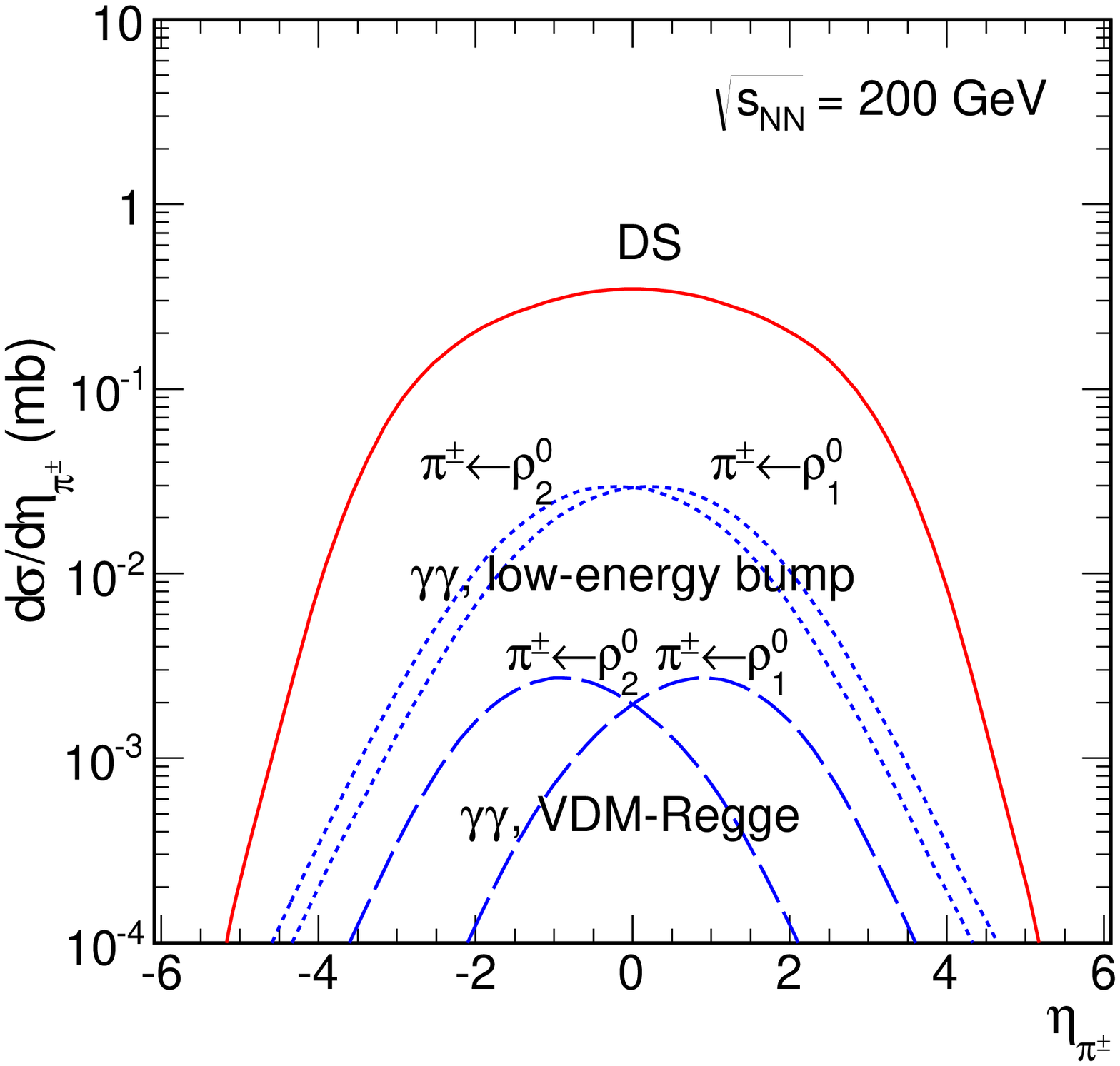}
   \caption{
\small Pseudorapidity distribution of charged pions
for double scattering (solid line), VDM-Regge photon-photon (dashed line)
and low-energy bump (dotted line) mechanisms 
for full phase space at $\sqrt{s_{NN}}$ = 200 GeV.
}
 \label{fig:dsig_detapi}
\end{figure}
%------------------------------------------------------------------------------

In Fig.~\ref{fig:dsig_detapi} we show distributions in pseudorapidity of 
the charged pions. The distributions extend over a broad range
of pseudorapidity. Both STAR collaboration at RHIC and the ALICE 
collaboration at LHC can observe only a small fraction of pions 
due to the rather limited angular (pseudorapidity) coverage $\eta \sim$ 0. 
While the CMS (pseudo)rapidity coverage is wider, 
it is not clear to us if the CMS collaboration has a relevant trigger 
to measure the exclusive nuclear processes described here.

%-----------------------------------------------------------------------------
\begin{figure}[!h]
\includegraphics[width=8cm]{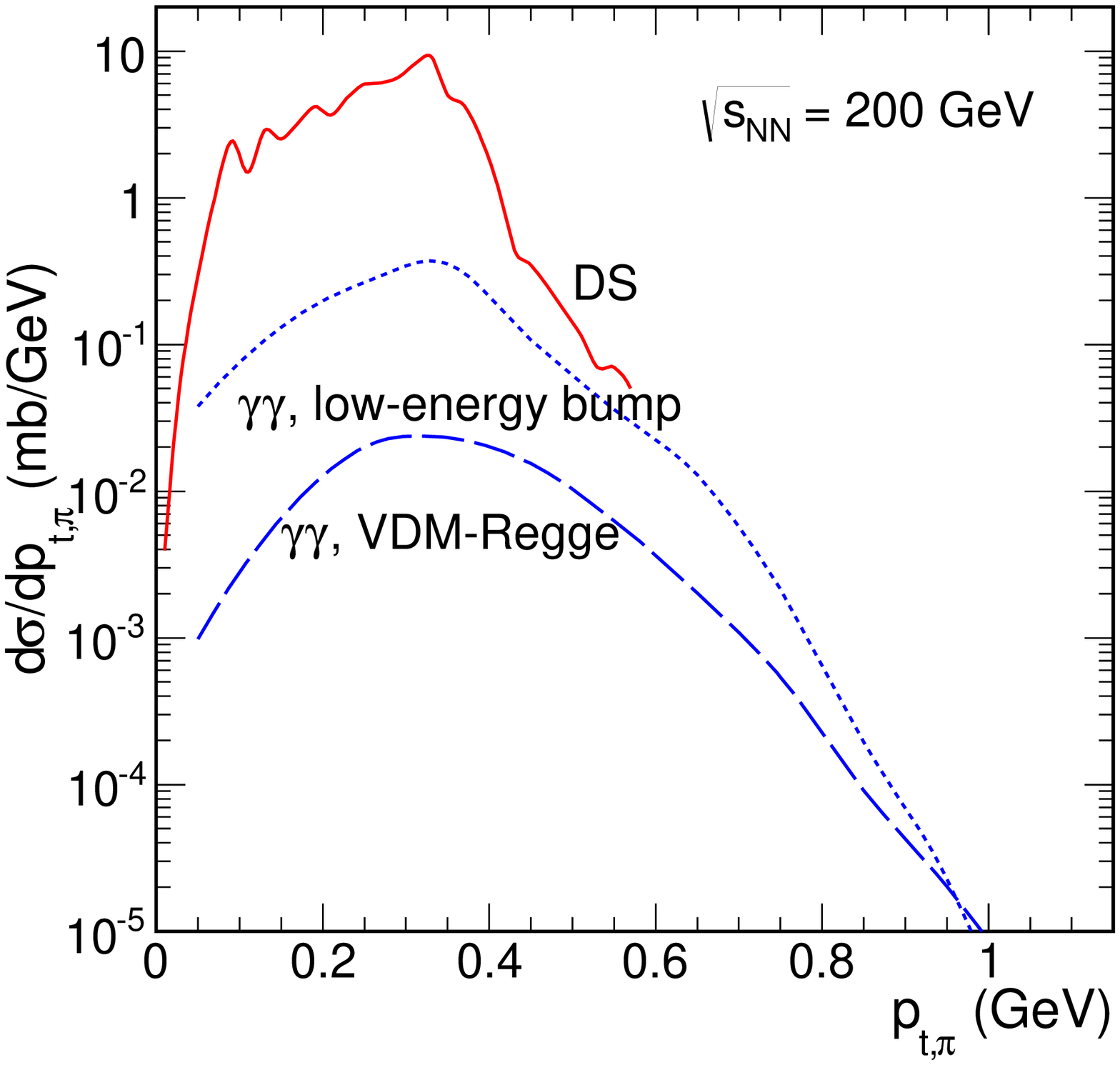}
   \caption{
\small Transverse momentum distribution of charged pions
for double scattering (solid line) low-energy bump (dotted line)
and VDM-Regge photon-photon (dashed line)
mechanisms for full phase space at $\sqrt{s_{NN}}$ = 200 GeV.
}
 \label{fig:dsig_dptpi}
\end{figure}
%------------------------------------------------------------------------------

For completeness in Fig.~\ref{fig:dsig_dptpi} we show distributions
in pion transverse momenta. 
%The distributions at RHIC and LHC have quite similar shapes. 
Since the $\rho^0$ mesons produced in the double-scattering
mechanism (photon-pomeron or pomeron-photon fusion) have very small 
transverse momenta, the transverse momenta of pions are limited
to $\sim m_{\rho^0}/2$. The distribution is relatively smooth,
as we have taken into account a smearing of $\rho^0$ meson masses. 
The sharp upper limit is an artifact of our maximal value of $\rho^0$
meson mass $m_{\rho}^{max}$ = 1.2 GeV. We have imposed this upper limit
because the spectral shape of ``$\rho^0$ meson'' above 
$m_{\rho} >$ 1.2 GeV is not under good theoretical control.
In principle, at larger $p_{t,\pi}$, the contribution coming from 
the decay of $\rho^0$ meson produced in photon-photon 
fusion can be larger as that of double scattering mechanism, as 
the transverse momentum of $\rho^0$ mesons are not strictly limited 
to small values. However, the cross section for such cases is expected 
to be rather small. Both STAR ($p_t >$ 0.1 GeV) and ALICE ($p_t >$ 0.1 GeV) 
experiments have a fairly good coverage in pion transverse momenta
and could measure such distributions.

%--------------------------
\section{Conclusions}
%--------------------------

In this work we have studied two-$\rho^0$  as well as four-pion
production in exclusive ultraperipheral heavy ion collisions,
concentrating on the double scattering mechanism of single $\rho^0$ production. 

Differential distributions for the two $\rho^0$ mesons, as well as
for four pions have been presented. The results (total cross section and differential
distributions) for the double scattering mechanism have been compared with 
the results for two-photon fusion discussed previously in the literature.
We have found that at the RHIC energy $\sqrt{s_{NN}}$ = 200 GeV 
the contribution of double scattering is almost two orders of magnitude 
larger than that for the photon-photon mechanism.

The produced $\rho^0$ mesons decay, with almost 100 \% probability, into 
charged pions giving large contribution to exclusive production of the
$\pi^+ \pi^- \pi^+ \pi^-$ final state.
We have made a comparison of four pion production via $\rho^0 \rho^0$
production (double scattering and photon-photon fusion) with experimental
data measured by the STAR collaboration for 
the $Au Au \to Au Au \pi^+ \pi^- \pi^+ \pi^-$ reaction. 
The theoretical predictions have a similar shape in four-pion invariant
mass as measured by the STAR collaboration (in a limited interval
of pion pseudorapidity and transverse momenta), but exhaust only about 
20 \% of the measured cross section. The missing contribution is probably
due to  the exclusive production of $\rho^0(1700)$ (excited state
of $\rho^0(770)$) resonance and its decay into four charged pions. 
We leave a theoretical calculation
for the latter mechanism for a separate analysis. We expect
that in the total phase space the contribution of double scattering
is similar to that for the $\rho^0(1700)$ resonant production.

At large $\rho^0 \rho^0$ (pseudo)rapidity 
separations and/or large $\pi^+ \pi^+$ ($\pi^- \pi^-$) (pseudo)rapidity
separations the double scattering contribution should dominate
over other contributions. The identification of the dominance region 
seems difficult, if not impossible, at RHIC.

The four charged pion final state is being analyzed by the ALICE
collaboration. We plan a separate careful analysis for the ALICE and/or
other LHC experiments. It would be interesting if different
mechanisms discussed in the present paper could be separated
and identified experimentally in the future. This requires, however, 
rather complicated correlation studies for four charged pions. 
Such a sudy will be presented elsewhere.

Similar double scattering mechanisms could be studied for different
vector meson production, e.g., for exclusive production of 
$\rho^0 J/\Psi$. Recently we have sudied
production of $J/\Psi J/\Psi$ pairs via two-photon mechanism in 
exclusive heavy ion reactions \cite{BCKSS2013}. A calculation
of the corresponding double-scattering contribution is very interesting
in the present context.

{\bf Acknowledgments}

We are indebted to Boris Grube for a discussion of the STAR 
experimental data and explanation of some details as well as providing
us numerical representation of the STAR experimental data.
The discussion with Daniel Tapia Takaki about four pion production
at ALICE (LHC) is kindly acknowledged. 
The careful reading of the manuscript by Christoph Mayer
is acknowledged.
This work was partially supported by 
N N202 236640 and 
N DEC-2011/01/B/ST2/04535.
A big part of the calculations within this analysis was carried out
with the help of the cloud computer system 
(Cracow Cloud One \footnote{cc1.ifj.edu.pl}) 
of Institute of Nuclear Physics (PAN).
%------------------------------------------------------------------------

\end{document}